\documentclass[a4paper,11pt]{article}
\usepackage{jheppub}
\usepackage{bm}
\usepackage{pgfplots}
\pgfplotsset{compat=newest}
\usetikzlibrary{shadows.blur}
\usepackage{float}

\def\beq{\begin{equation}}
\def\eeq{\end{equation}}
\def\bea{\begin{eqnarray}}
\def\eea{\end{eqnarray}}

\def\eq#1{{Eq.~(\ref{#1})}}

\renewcommand{\v}[1]{ \ensuremath{ {\bm{#1}} }}                  
\parskip=\itemsep               

\title{ Born-Oppenheimer Renormalization group for High Energy Scattering: the Setup and the Wave Function.}
\author[a]{Haowu Duan,}
\author[a,b]{Alex Kovner}
\author[c]{and Michael Lublinsky}

\affiliation[a]{Physics Department, University of Connecticut, 2152 Hillside Road, Storrs, CT 06269, USA}
\affiliation[b]{ExtreMe Matter Institute EMMI,
GSI Helmholtzzentrum fuer Schwerionenforschung GmbH,
Planckstrasse 1,
64291 Darmstadt,
Germany}
\affiliation[c]{Physics Department, Ben-Gurion University of the Negev, Beer Sheva 84105, Israel}

\abstract{We develop an approach to QCD evolution based on the sequential Born-Oppenheimer approximations  that include higher and higher frequency modes as the evolution parameter is increased. This Born-Oppenheimer renormalization group is a general approach which is valid for the high energy evolution as well as the evolution in transverse resolution scale $Q^2$. In the former case it yields the frequency ordered formulation of high energy evolution, which includes both the eikonal splittings which produce gluons with low longitudinal momentum,  and the DGLAP-like splittings which produce partons with high transverse momentum. In this, first paper of the series we lay out the formulation of the approach, and derive the expression for the evolved wave function of a hadronic state. We also discuss the form of the $S$-matrix which is consistent with the frequency ordering. }

\begin{document}
\maketitle

\pagestyle{empty}

\mbox{}

\pagestyle{plain}

\setcounter{page}{1}
\author{}

\abstract{ }
\dedicated{}
\preprint{}

\date{\today}

\section{Introduction }

The subject of evolution of hadronic scattering amplitudes at high energy has been at the forefront of QCD research for many years. A lot is known about this evolution. The evolution equation within a perturbative realm, the BFKL equation \cite{bfkl,bfkl2,bfkl3} was derived many years ago, and is known at next to leading order (NLO) \cite{NLOBFKL1,NLOBFKL2}. 
It has been generalized to account for multiple scattering effects on dense targets. Here several approaches were pursued which lead to different formulations of the same physical process: the so called Balitsky hierarchy \cite{Bal1,Bal2, Bal3}, the Balitsky-Kovchegov (BK) equation \cite{KOV} and the functional JIMWLK equation \cite{jimwlk1,jimwlk2,jimwlk3,jimwlk4,jimwlk5,jimwlk6,cgc1,cgc2,cgc3} (for NLO expression see \cite{BC,BClast,Grab,nlojimwlk1,nlojimwlk2,LuMu}).
In this paper we will be using the language of JIMWLK approach. 

The original and subsequent formulations of the high energy evolution employed the $k^+$ ordering, where as one evolves the wave function of a right moving hadron towards higher energies, new gluon emissions are taken into account and those emissions are ordered in the decreasing order in longitudinal momentum $k^+$.
At NLO all the evolution equations are known to suffer from instability due to large corrections. In the context of BFKL this instability leads  in some cases to negative cross section, and the problems are similar in the BK and JIMWLK frameworks. It has been recognized that the origin of the large corrections is the fact that at leading order (LO) one assumes that the evolution is dominated by emission of gluons with roughly the same transverse momenta as that of the parent emitters.  The energy evolution is then essentially just the evolution in the longitudinal momentum of the emitted gluons. This turns out to be not quite true, and it is the emission of gluons with very large transverse momenta that is unrestricted in LO JIMWLK that leads to the said instability. The prevailing wisdom is that large perturbative corrections have to be resummed  in order to obtain a stable and sensible evolution. It has been recognized long ago that  unrestricted emission of high momentum  gluons is allowed due to  non-conservation of energy in the leading order BFKL and JIMWLK equations.

 To rectify this problem it was proposed to impose the so called "kinematical constraint" \cite{Kutak,Motyka,Beuf, Vera:2005jt, Iancu:2015vea, Iancu:2015joa,Ducloue:2019ezk}  which in effect imposes double ordering in the evolution -  both in $k^+$ and $k^-$. Several  approaches to impose the constraint have been proposed. The drawback of these approaches, as we see it, is that they are rather "ad hoc" especially in the case on nonlinear evolution, in the sense that they do not provide a comprehensive physical picture of the situation, and therefore leave questions as to preference of one or another prescription unanswered.

Apart from the energy non-conservation there is another set of large corrections at NLO JIMWLK which involve large transverse logarithms. These corrections are related to the DGLAP evolution within JIMWLK Hamiltonian. The problem here is that JIMWLK splitting (just like BFKL) occur with the approximate form of the splitting function, which is taken to be equal to the low $x$ limiting form over the whole range of $x$. This leads to significant overestimate of the speed of the evolution. Restoring the correct form of the splitting function must slow down the evolution considerably and also allow for a possibility to use JIMWLK-like evolution to bridge between the extreme high energy limit and lower energies where the nonlinear effects may still be important. This is of course very important for applications at the EIC. The way to resum the DGLAP logarithms within the JIMWLK approach was formulated in \cite{ourdglap}.

Our goal in this series of  papers is to develop the "first principles" physical approach which naturally addresses and rectifies both problems. In this approach we rely on the physical picture discussed in \cite{armesto}, and  later in \cite{Ducloue:2019ezk} which stresses that the evolution parameter most appropriate for the energy evolution is the frequency, $k^-$, and not the longitudinal momentum $k^+$, of the gluon modes. 
The importance of frequency evolution was recognized already in the context of the linear BFKL equation at NLO \cite{gavin1,gavin2,gavin3,gavin4}.
As the energy of scattering increases, modes with higher frequency "freeze" for long enough in order to have time to scatter coherently on a given target. Thus evolution in energy is achieved by including modes with higher frequency in the wave function of a fast moving hadron. 
As we will see, the frequency evolution is not just more physical in the context of increasing energy of scattering, but it encompasses the standard QCD DGLAP evolution. This is quite natural, since the DGLAP splitting process  increases the transverse momentum of emitted gluons  and thus produces gluons with higher frequency than their parents. Thus the frequency evolution naturally unifies both the BFKL type and DGLAP type physics within the same framework and should be fundamental for unifying the low x and intermediate x physics, the challenge that has been taken up by the community in recent years \cite{balitsky-tarasov1,balitsky-tarasov2,balitsky-tarasov3,lowix1,lowix2,lowix3,lowix4}.

 Another feature which is conceptually very satisfactory, and which is the basis of our approach, is that frequency evolution physically can be formulated as the sequence of Born-Oppenheimer (BO) approximations. In  every step one includes into the wave function modes with higher frequency. Their wave function needs to be calculated given a fixed background of slower "valence" modes.  This procedure recurs until the highest frequency relevant for a particular physical observable is reached.  We dub this procedure the "Born-Oppenheimer renormalization group" approach. This procedure implements directly the frequency ordering into the QCD evolution. There is no need to implement the  longitudinal momentum ordering in this approach. However a form of longitudinal momentum ordering arises effectively  in the calculation through the use of eikonal emission vertices for those modes that have vastly different longitudinal momenta. Importantly though, we do not impose the eikonal emission on all emission vertices.  Quite to the contrary we also consider interactions between gluons with similar longitudinal momenta, as long as their frequencies are very disparate. These interactions are not negligible within the context of $k^-$ ordered evolution. Instead they account for DGLAP like emissions which are also included in our calculation on par with the BFKL like emissions initiated by the eikonal vertices.

 In this, first  paper of the series we formulate the Born-Oppenheimer renormalization group approach to calculation of  the evolution of hadronic light cone wave function (LCWF) at leading perturbative order. This amounts to calculating LCWF of faster gluons in the slow mode background perturbatively. In this paper we do not consider effects of high partonic density in the projectile wave function, and treat the projectile as dilute \cite{Kovner:2005nq}.  The generalization to dense projectile is left to future work. 
 
Also in this paper we do not include quarks in the calculation.  Quarks are not negligible in the $k^-$ ordered approach even in the leading order, since the Born-Oppenheimer approach includes DGLAP type physics on par with the BFKL like emissions. However since our main goal here is to present the basic ideas of the Born-Oppenheimer approach, in the interest of simplicity of the presentation we defer inclusion of quarks to future publications.
 
 This paper is structured as follows.
 In Sec. 2 we set out the basic approach. We analyze the interaction Hamiltonian $H_{SF}$
 between the fast and slow modes and show explicitly that both the eikonal and transverse momentum splitting vertices are important for this interaction. Using this interaction Hamiltonian we derive perturbatively the LCWF of a hadronic system evolved from some initial frequency $E_0$ to frequency $E$, which will serve as the basis of the calculations of the evolution in the subsequent work.
 
\begin{figure}[H] 
\centering 
\begin{tikzpicture}
\begin{axis}[scale=1.2,
        axis lines=left,             
        axis line style={-latex},
	 xmin=0,   xmax=4,
	 ymin=0,   ymax=2.5,
	 xtick=\empty,
	 ytick=\empty,
	 xlabel={$k^-$}, 
         ylabel={$\Psi$},
          x label style={at={(axis description cs:1,0)}}, 
          y label style={rotate=270, at={(axis description cs:0,1)}},
	clip=false,
	]    
	 \addplot[domain=0:2.5, dash pattern=on 5pt off 3pt] ({2},{x}); 
	 \addplot[domain=0:2.5, dash pattern=on 5pt off 3pt] ({2.8},{x});
	 \node at (axis description cs:0.5,-0.1) [anchor=south] {$E$};
	 \node at (axis description cs:0.7,-0.1) [anchor=south] {$Ee^\Delta$};
	   \addplot[domain=0:2.8, smooth, thick, blue, mark=none] {2*(x/2.8)^2 * (2.8-x) * sin(pi*x/2.8)*500 / (1+x)^2} 
    node[pos=0.4, above, font=\small, xshift=-18pt] {\textcolor{blue}{Projectile}};
 \addplot[domain=2.8:4, smooth, thick, red, mark=none] {(5)*exp(-20*(x-2.8))*(20*x-56)} 
    node[pos=0.2, left, yshift=5pt, font=\small, xshift=45pt] {\textcolor{red}{Target}};    
      {\color{red}
    \draw[decorate,decoration={brace,amplitude=5pt,mirror},very thick] (0,0) -- (2,0) node[midway,below=8pt,font=\small] {Slow/ Valence};
    \draw[decorate,decoration={brace,amplitude=5pt,mirror},very thick] (2,0) -- (2.8,0) node[midway,below=8pt,font=\small] {Fast};
    }
	\end{axis}
\end{tikzpicture}
\caption{A qualitative picture of the projectile and target wave functions in the $k^-$ ordering scheme.}
\label{Fig} 
\end{figure}
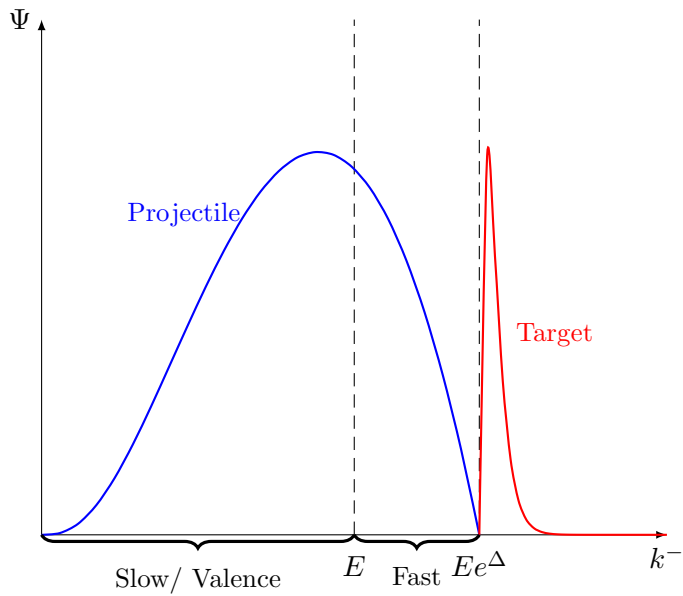

In Sec. 3 we derive the operator of the total scattering matrix ($S$-matrix), as the simplest interesting observable. 
The $S$-matrix operator $\hat S$ is a quantum operator on the Hilbert space which is a product of the Hilbert spaces of the projectile ($P$) and the target ($T$). The projectile is taken to have the total longitudinal momentum $P^+$ while the central mass collision energy squared is $s$.
The projectile and target  degrees of freedom are separated by the frequency scale $E$, as in the calculation of the evolution (see Fig \ref{Fig}). The use of the very same separation scale is inherent in our approach. The initial state of the projectile plus target system before the 
scattering has the form
 \begin{equation}\label{in}
 |\Psi_{in}\rangle=|\Psi_P\rangle\otimes|\Psi_T\rangle
 \end{equation}
 where $\Psi_P$ is the wave function of the projectile and is a state in the Hilbert space of modes with frequencies smaller than the separation scale, $k^-<E$, while $\Psi_T$ is the wave function of the target and is a state in the complementary part of the Hilbert space, $k^->E$.  
Our calculation of the $S$-matrix is valid for targets  with very sharp  distribution of frequencies around $k^-\simeq E$. For such targets the total collision energy is 
$s=2P^+E$. Fixing $P^+$ therefore fixes the Lorentz frame in which the calculation is performed.

 The interaction Hamiltonian between the projectile and target modes, $H_{PT}$ is intimately related to the interaction Hamiltonian between the fast and slow modes in the projectile. The $S$-matrix operator is computed as the time evolution operator
 \begin{equation}\label{S}
 \hat S=\lim_{\tau\rightarrow\infty}{\cal T}\exp\left\{i\int_0^\tau dx^+ H^I_{PT}(x^+)\right\}
 \end{equation}
where $H^I_{PT}$ is the interaction picture Hamiltonian of the interaction. 
 The form of the $S$-matrix is therefore intimately related to the form of the operator that diagonalizes the Hamiltonian. It is affected by the approximations we employ for the diagonalizing operator, or equivalently for the evolved wave function, as the consistently between the $S$-matrix and the diagonalizing operator has to be preserved in any consistent approximation.  For explicit calculations we limit ourselves  to the situation where the target fields are small ("dilute target"). In this situation we find the explicit form of the $S$-matrix as an operator on the projectile Hilbert space.
 
 In Sec. 4 we use the LCWF obtained in Sec.2 to make connection with the derivation of NLO JIMWLK Hamiltonian in \cite{LuMu}.
We  review/discuss the way in which  large transverse logarithms arise in \cite{LuMu} with the emphasis  on the one gluon amplitude computed at NLO, and point out some subtleties that arise when trying to relate this calculation to JIMWLK Hamiltonian.
We then use  the evolved wave function $|\Psi_P\rangle$ derived in Sec 2. to compute the same amplitude to order $g^3$.
We demonstrate that the calculation using the Born-Oppenheimer wave function  expanded to order $g^3$ reproduces the large transverse logarithms  calculated in \cite{LuMu}.
 
 Finally, we conclude in Sec. 5 with a short discussion.

\section{The Born-Oppenheimer approximation for QCD evolution}

In \cite{armesto} we have argued that in the context of high energy evolution the frequency of the fastest relevant modes in the wave function is determined by the "Ioffe time cutoff", $\tau\equiv 1/E$.  This is the time a given fluctuation should exist in order to scatter coherently off the target,  and it is thus related to the physical longitudinal extent of the target. 
When the energy is increased by boosting the projectile with boost parameter $\Delta$, the modes with frequencies just above those determined by the Ioffe time cutoff, $E<k^-<Ee^{\Delta}$ become important due to time dilation (or Lorentz contraction if the energy is imparted to the target) and participate in coherent scattering.

In fact the utility of the mode separation based on frequency (we will refer to the logarithm of frequency as "rapidity" in the following) is more general. Even if the scattering is not fully coherent, but is due to scattering with high transverse momentum transfer, the projectile gluon has to exist long enough to be resolved by the target field. In the case of the high momentum transfer this life time is determined by the interaction time which is inversely proportional to the square of the momentum transfer. The actual value of this cutoff is unimportant for our derivation. What is important however, is that this cutoff provides natural separation between the fast and slow modes. 

To obtain the frequency evolution of physical observables one therefore needs to calculate the wave function of the fast modes (those that live a little above the Ioffe time cutoff) in the presence of slower, lower frequency modes. This is the classic situation suitable for the Born-Oppenheimer approximation in which, in order to solve for the wave function of the fast degrees of freedom, one freezes the dynamics of the slow modes and treats them as a fixed, time independent background. Our idea in this paper is to apply this approximation to the calculation of the QCD LCWF.

We start with formulating  the Born-Oppenheimer approximation in the context of hadronic evolution. As in the standard BO approximation in quantum mechanics, we partition  the projectile Hilbert space  into fast ($F$) and slow ($S$). The separation is based on the frequency $k^-$ of the appropriate modes in the absence of interaction 
(Fig.\ref{Fig}). The dynamics of the slow modes is described by the Hamiltonian $H_S$, which includes the kinetic term $H_0$ and self-interactions. Similarly, for the fast modes we have  $H_F$.  The two sets of modes interact via an interaction Hamiltonian $H_{SF}$,  so that the total Hamiltonian of the system is $H_S+H_F+H_{SF}$.
Since we assume the coupling to be weak, such a separation of modes is sensible. 

We  assume that  the LCWF  $|\psi_0\rangle_S$ of the slow modes below some frequency cutoff $E$ is known. In the absence of interaction between the slow and fast modes,  the vacuum of the fast modes is just the Fock vacuum $|0\rangle_F$. 
The total wave function is thus a direct product $|0\rangle_F \otimes |\psi_0\rangle_S$.

Treating $H_{SF}$ as a weak perturbation one can  find the ground 
state of the fast modes perturbatively. More generally, one can find perturbatively the operator that diagonalizes the Hamiltonian of the coupled system. Neglecting (or in the perfect world solving for) the self interaction of the fast and soft modes, the time evolution of the coupled system is  given by the interaction picture evolution operator 
 \begin{equation}\label{usf}
 U(0,\tau)={\cal T}\exp\{i\int_0^\tau dx^+ H^I_{SF}(x^+)\}
 \end{equation}
  where the  Hamiltonian $H^I_{SF}$ is expressed in terms of the interaction picture fields.   
 According to Low's theorem, the  operator diagonalizing the interaction is given by 
 \begin{equation}\label{omega}
 \Omega=\lim_{\tau\rightarrow\infty}U(0,\tau)
 \end{equation}
 Thus the ground state wave function of the coupled system is given by
 \begin{equation}
 |\Psi_P\rangle=\Omega\,|0\rangle_F \otimes |\psi_0\rangle_S
 \end{equation}
 We will find this relation between the wave function and the evolution operator useful also when discussing the form of the scattering matrix in Sec. 3.

To apply the BO approximation our first goal is to map out the interaction Hamiltonian on the light front.

\subsection{The light cone  Hamiltonian -- a recap}
This subsection is a recap of the light front Hamiltonian formalism which we use throughout this paper.
We start with the QCD Lagrangian density in the light cone coordinates for the projectile degrees of freedom
  \begin{equation}\label{LYM}
      \mathcal{L}_{YM}=-\frac{1}{4} F^{\mu\nu}_a F_{\mu\nu}^a=\frac{1}{2} (F^{+-})^2+F^{+i}F^{-i}-\frac{1}{4}F^{ij}F_{ij}
      \end{equation}
      with
      \begin{eqnarray}
      &&F^{+-}=\partial^+A^--D^-A^+;\ \qquad  \ \ F^{+i}=\partial^+A^i-D^iA^+\\
      &&F^{-i}=\partial^-A^i-D^iA^-; \ \qquad \ \ F^{ij}=\partial^iA^j-D^jA^i\nonumber
      \end{eqnarray}
      where  $D^{ab}A^b=\partial A^a-gf^{abc}A^bA^c$.
      We are working in the projectile light cone gauge $A^+=0$. As usual, before setting the $A^+$ to zero in the Lagrangian we have to derive the "equation of motion" that follows from differentiating the Lagrangian with respect to $A^+$. 
  This equation of motion reads (in $A^+=0$ gauge)
  \begin{equation}\label{+con}
  D^-F^{+-}+D^iF^{-i}=D^-\partial^+A^-+D^i[\partial^-A^i-D^iA^-]=\partial^-\partial^+A^--\partial^2A^-+\partial^i\partial^-A^i+O(g)=0
  \end{equation}
  where we have made explicit the terms not involving the coupling constant.
  The field $A^-$ in the Lagrangian is non-dynamical as it has no kinetic term. Hence it can be eliminated via equation of motion
  (see (\ref{-con})).
  
 The Hamiltonian for  the projectile gluons is given by the Legendre transform of the Lagrangian with the conjugate momenta
 \begin{equation}
 { \Pi}^a_i(x)=\partial^+ A^a_i(x)
 \end{equation}
The result of the Legendre transform is the  light front QCD Hamiltonian:
\begin{equation}
H=\frac{1}{2}\int_{x^-,\v x}\Big[\Pi^a(x^-,\v x)\Pi^a(x^-,\v x)+B(x^-,\v x)B(x^-,\v x)\Big]
\end{equation}
where the longitudinal electric field $\Pi^a$ and the transverse magnetic field $B^a$ are given by
\begin{eqnarray}
\Pi^a(x^-,\v x)&\equiv&F^{+-}_a(x^-,\v x)=\frac{1}{\partial^+}D_i^{ab}\partial^+A_i^b=\partial_iA^a_i-g\frac{1}{\partial^+}f^{abc}A_i^b\partial^+A_i^c\\
B^a(x^-,\v x)&=&\epsilon_{ij}\Big[\partial_iA_j^a-\frac{g}{2}f^{abc}A_i^bA_j^c\Big]
\end{eqnarray}
Decomposing the fields in Fourier components\footnote{We denote by boldface letters the transverse components of coordinate and momentum vectors.}
 \begin{equation}\label{Ak}
 A^i_a(x^-,\v x)=\int \frac{d^3k}{(2\pi)^3}\left[A^i_a(k^+,\v k)e^{-i k\cdot  x}+A^{\dagger i}_a(k^+, \v k)e^{i k\cdot x}\right]
 \end{equation}
  where $k\cdot x\equiv k^+x^-- \v k\cdot \v x$, we Fourier transform $\Pi^a$  
\begin{eqnarray}\label{pif}
\Pi^a(p^+,\v p)&=&-i\v p_iA^a_i(p^+,\v p)+gf^{abc}\int_{k^+>0}A_i^b(p^++k^+,\v p+\v k)\frac{2k^++p^+}{p^+}A_i^{c\dagger}(k^+,\v k)\nonumber\\
&+&gf^{abc}\int_{p^+>k^+>0}A_i^b(k^+,\v k)\frac{k^+}{p^+}A_i^c(p^+-k^+,\v p-\v k).
\end{eqnarray}
Here  we are using the shorthand notation $\int_k \equiv \int \frac{d^3k}{(2\pi)^3}$. Also $A^a_i(k)\equiv\frac{1}{\sqrt{2k^+}}\hat a^a_i(k); \ \ \  A^{a\dagger}_i(k)\equiv\frac{1}{\sqrt{2k^+}}\hat a^{a\dagger}_i(k)$, with $\hat a^\dagger$ and $\hat a$ - creation and annihilation operators.
The fields satisfy the canonical light cone commutation relation
\begin{equation}\label{comm}
[A^a_i(p),A^{b\dagger}_j(q)]=(2\pi)^3\delta^{ab}_{ij}\frac{1}{2p^+}\delta^3(p-q)
\end{equation}
Since $\Pi^a$ is real in coordinate space, the negative $p^+$ components are not independent:
\begin{eqnarray}
\Pi^a(-p^+,-\v p)=\Pi^{a\dagger}(p^+,\v p)
\end{eqnarray}
For the magnetic field we get analogously
\begin{eqnarray}
B^a(p^+,\v p)&=&\epsilon_{ij}\Big[-i\v p _iA^a_j(p^+,\v p)-gf^{abc}\int_{k^+}A_i^b(k^++p^+,\v k+\v p)A_j^{c\dagger}(k^+,\v k) \nonumber \\
&-&\frac{g}{2}f^{abc}\int_{p^+>k^+>0}A_i^b(p^+-k^+,\v p-\v k)A_j^c(k^+,\v k)\Big]
\end{eqnarray}
\begin{eqnarray}
B^a(-p^+,-\v p)&=&\epsilon_{ij}\Big[i\v p_iA^{a\dagger}_j(p^+,\v p)-gf^{abc}\int_{k^+}A_i^{b\dagger}(k^++p^+,\v k+\v p)A_j^{c}(k^+,\v k)\nonumber \\
&-&\frac{g}{2}f^{abc}\int_{p^+>k^+>0}A_i^{b\dagger}(p^+-k^+,\v p-\v k)A_j^{c\dagger}(k^+,\v k)\Big]
\end{eqnarray}
The Hamiltonian in momentum space reads
\begin{equation}
H={1\over 2} \int_p\left[ \Pi^a(p^+,\v p)\Pi^{\dagger a}(p^+,\v p)+B^a(p^+,\v p)B^{\dagger a}(p^+,\v p)\right]
\end{equation}

\subsection{The coupling between fast and slow modes - a rough guide.}

Our first goal is to determine the coupling between the slow 
and fast modes.

As mentioned in the Introduction,
 $E$ is the maximal frequency associated with the projectile. More accurately,
we define a narrow band of frequencies $E<p^-\equiv\frac{\v p^2}{2p^+}<Ee^\Delta$, where $\Delta$ is small. Eventually we will take it to be infinitesimally small in order to derive the differential form of the evolution. The fast modes are those that populate the above frequency interval, or "window" as we will sometimes refer to it. The slow degrees of freedom are field modes with frequencies  well below $E$. 
The distinction between "slow" and "fast" is of course dynamical since the value of  $E$ changes with evolution.

At this point all we really need is to extract that part of QCD interaction vertex which couples modes with very different frequencies.   The relevant vertex is the triple gluon vertex, although the four gluon vertex will also become important in higher order calculations. For book keeping purposes any component of the gauge field $A^a_i(k)$ can be split into the low and high frequency parts
\begin{equation}
A^a_i(k)=A^a_{Si}(k)\,\theta\left(E-\frac{\v k^2}{2k^+}\right)+A^a_{Fi}(k)\,\theta\left(\frac{\v k^2}{2k^+}-E\right)
\end{equation}
The purpose of this splitting at the moment for us is simply to keep track which frequency is high and which is low in the interaction vertex.

In this subsection our discussion will be a little loose and semi intuitive. It is meant to make clear what physical effects are taken into account in the Born-Oppenheimer approximation approach. We will put it on a firm mathematical basis in the next subsection. 

We start by noting, that there are two physically distinct regions in the phase space which are populated by  modes which are  fast relative to any given mode $(k^+,\v k)$. The first region is "BFKL"-like, i.e.  $(p^+,\v p)$ with $p^+\ll k^+$ and $\v p^2\sim \v k^2$. 
The second region is "DGLAP"-like with $p^+\sim k^+$ and $\v p^2\gg \v k^2$. Let us introduce  compact notations for the phase space
\begin{equation}
\Omega_{BFKL}:\, p^->k^-;\,\, p^+\ll k^+;\,\,\v p^2\sim \v k^2; \qquad
\Omega_{DGLAP}:\, p^->k^-;\,\, p^+\sim k^+;\,\, \v p^2\gg \v k^2
\end{equation}
These two regions have a nontrivial overlap which is the doubly logarithmic part of phase space. One needs to separate between the two in a clean way to avoid any double counting. We will do this in the next subsection, but meanwhile will proceed with naive discussion without worrying too much about the overlap.

 The Hamiltonian density of a given fast mode  with momentum $p$, where $p^-$ is a momentum in the "window" is
 \begin{equation}\label{Hp}
{\cal H}(p)={1\over 2} \left[ \Pi_F^a(p^+,\v p)\Pi_F^{\dagger a}(p^+,\v p)+B_F^a(p^+,\v p)B_F^{\dagger a}(p^+,\v p)\right]
\end{equation}
 where  the fast components of the electric  and magnetic fields are denoted as $\Pi^a_F(p)$ and $B^a_F(p)$. 
 
 Since $\Pi_F$ and $B_F$ are nonlinear functions of the vector potential, they depend on the slow modes as well.
 The contribution of the low frequency modes to $\Pi^a_F(p)$ ad $B^a_F(p)$ is of two distinct types.
 The first contribution is due to $A^a_{Si}(k^+,\v k)$ such that $k^+\gg p^+; \ \ \ \v k\sim \v p$. Those modes 
 create a color charge density which serves as a source for $A_F(p)$
\begin{eqnarray}\label{rho}
\rho^a(p^+,\v p)=&&-i
f^{abc}\int_{k\in\Omega_{BFKL}} (2k^++p^+)A^{b\dagger}_{Sj}(k^++p^+,\v k+\v p) A^c_{Sj}(k^+,\v k)\nonumber\\
&&-i
f^{abc}\int_{k\in\Omega_{DGLAP}} (2k^++p^+)A^{b\dagger}_{Sj}(k^++p^+,\v k+\v p) A^c_{Sj}(k^+,\v k)\nonumber\\
&&\approx -2i
f^{abc}\int_{k\in \Omega_{BFKL}}k^+A^{b\dagger}_{Sj}(k^+,\v k+\v p) A^c_{Sj}(k^+,\v k)
\end{eqnarray}
In the Hamiltonian (\ref{Hp}), this  contributes $"\rho A_F"$ term - the usual eikonal coupling of the color charge density to the fast mode,
modified only to account for hierarchy of frequencies of the modes that couple. The interaction between this charge density and the fast gauge field is the basis of the derivation of the BFKL evolution as well as JIMWLK evolution. The longitudinal momenta in \eqref{rho} are strongly ordered since only in this case the QCD vertex reduces to the eikonal vertex.

In \eqref{rho} we have explicitly indicated  that the charge density is restricted to contribution of the modes with low frequency only. Thus, for example modes with too high transverse momentum are not included in the color charge density, even if their longitudinal momentum is much higher than that of the field mode they couple to. It is not really necessary to explicitly show the bound on the frequency in the operator, since in the actual calculation,  the operator $\rho^a$ acts on the projectile state which does not contain any particles with the frequency higher than 
$p^-\sim E$. Thus $\rho^a$ will  annihilate such a state even if it formally contains gluon operators with higher frequencies. Still we find it convenient to keep the restriction on the frequency in the operator itself, at least for the purpose of the present discussion.
 
 We note that  in principle there is also a magnetic field created by the slow modes
\begin{eqnarray}
b^a(p^+,\v p)=&&-g\epsilon_{ij}f^{abc}\int_{k\in\Omega_{BFKL}}A_{Si}^{b\dagger}(k^++p^+,\v k+\v p)A_{Sj}^{c}(k^+,\v k)\\
&&-g\epsilon_{ij}f^{abc}\int_{k\in\Omega_{DGLAP}}A_{Si}^{b\dagger}(k^++p^+,\v k+\v p)A_{Sj}^{c}(k^+,\v k)\nonumber
\end{eqnarray} 
 However, as is well known the eikonal type magnetic field interaction is suppressed relative to the electric one,  since it does not contain and enhancing fact or the type $k^+/p^+$. The background magnetic field will not play an important role in our discussion.

Another type of the interaction comes from having one slow field in the quadratic term in $\Pi^a_F(p)$ and/or $B^a_F(p)$. 
The first term in eq.\eqref{pif} then reduces to
\begin{eqnarray}
&&gf^{abc}\int_{k^+>0}A_i^b(p^++k^+,\v p+\v k)\frac{2k^++p^+}{p^+}A_i^{c\dagger}(k^+,\v k)\rightarrow\\
&&gf^{abc}\left[\int_{k\in\Omega_{DGLAP}}A_{Si}^b(k^+,\v k)\frac{2k^+-p^+}{p^+}A_{Fi}^{c\dagger}(k^+-p^+,\v k-\v p)+
A_{Fi}^b(p^++k^+,\v p+\v k)\frac{2k^++p^+}{p^+}A_{Si}^{c\dagger}(k^+,\v k)\right]\nonumber\\
&&+gf^{abc}\int_{p^+\gg k^+}A_{Fi}^b(p^++k^+,\v p+\v k)\frac{2k^++p^+}{p^+}A_{Si}^{c\dagger}(k^+,\v k)\nonumber
\end{eqnarray}
The last term here is an interesting one. It involves the slow field with very small longitudinal momentum. This obviously mean that its transverse momentum is also very small so as to keep its frequency relatively low. The interaction with this field is neither what we would naturally call "BFKL" or "DGLAP" type. The gluons created by this mode are similar to what in the terminology of SCET are called "soft gluons". To underscore the different nature of these fields we will reserve a special notation for them and will denote them by $\alpha_i^a$. 

We do, however  have to be aware of the fact that these gluons are not universally "soft". They are "soft" relative to the other gluons with which they interact, however their softness does not mean that all their momenta are of order $\Lambda_{QCD}$, as for the genuine soft gluons of SCET \cite{SCET}. In particular the creation operator $\alpha_i^{\dagger a}$ creates a bona fide gluon state, and we can associate physical probability with such a state. 
Interestingly, analyzing the large logarithmic contributions to NLO JIMWLK \cite{LuMu}, one finds that some of these large logarithms arise due to interactions involving $\alpha_i^a$. The culprit is  the amplitude of  one gluon component in the  vacuum wave functional at NLO, where this gluon is created by $\alpha_i^{\dagger a}$.  We will come back to this point in more detail in Section IV. However in the bulk of the present paper  we will be dealing with the leading order calculation only. To this order the soft fields are unimportant. We keep their contributions for now, but will drop them very soon so as not to clutter notations. Nevertheless it is important to keep in mind that at higher orders one needs to include the contributions of these gluons into physical amplitudes, and therefore have a fully consistent treatment of these soft gluons. 

The other term in $\Pi_F^a(p)$ eq.\eqref{pif} can be written as 
\begin{eqnarray}
&f^{abc}\int_{p^+>k^+>0}A_i^b(k^+,\v k)\frac{k^+}{p^+}A_i^c(p^+-k^+,\v p-\v k)\rightarrow
f^{abc}\int_{k\in\Omega_{DGLAP}}A_{Si}^b(k^+,\v k)\frac{2k^+-p^+}{p^+}A_{Fi}^c(p^+-k^+,\v p-\v k) \nonumber
\\& +f^{abc}\int_{p^+\gg k^+}A_{Si}^b(k^+,\v k)\frac{2k^+-p^+}{p^+}A_{Fi}^c(p^+-k^+,\v p-\v k)
\end{eqnarray}
The last term again  contains the soft field $\alpha_i^b(k)$.

We can now write down the fast electric field, including the interaction between fast and slow modes as
\begin{eqnarray}
\Pi_F^a(p)&&=-i\v p_iA^a_{Fi}(p^+,\v p)-2g
f^{abc}\int_{k\in\Omega_{BFKL}}\frac{k^+}{p^+}A^{b\dagger}_{Sj}(p^+,\v k-\v p) A^c_{Sj}(k^+,\v k)\\
&&\hspace{-0.5cm}+gf^{abc} \Big\{\int_{p^+\gg k^+;\v k^2\sim \v p^2}
\left[A_{Fi}^b(p^+,\v p+\v k)\alpha_{i}^{c\dagger}(k^+,\v k)+A_{Fi}^b(p^+,\v p+\v k)\alpha_{i}^c(k^+,-\v k)\right]
\nonumber\\
&&\hspace{-0.5cm}+\left[\int_{k\in\Omega_{DGLAP}}A_{Si}^b(k^+,\v k)\frac{2k^+-p^+}{p^+}A_{Fi}^{c\dagger}(k^+-p^+,\v k-\v p)
+
A_{Fi}^b(p^++k^+,\v p+\v k)\frac{2k^++p^+}{p^+}A_{Si}^{c\dagger}(k^+,\v k)\right]\nonumber\\
&&+\int_{k\in\Omega_{DGLAP}}A_{Si}^b(k^+,\v k)\frac{2k^+-p^+}{p^+}A_{Fi}^c(p^+-k^+,\v p-\v k)\Big\}\nonumber
\end{eqnarray}
where we have neglected $k^+$ relative to $p^+$ in the first two lines.

The same exercise for the magnetic field yields
\begin{eqnarray}
&&\epsilon_{ij}gf^{abc}\int_{k^+}A_i^b(k^++p^+,\v k+\v p)A_j^{c\dagger}(k^+,\v k)\rightarrow\\
&&~~~~~\epsilon_{ij}gf^{abc}\int_{k\in\Omega_{DGLAP}}\left[A_{Fi}^b(k^++p^+,\v k+\v p)A_{Sj}^{c\dagger}(k^+,\v k)+A_{Fi}^{b\dagger}(k^+-p^+,-\v k-\v p)A_{Sj}^c(k^+,-\v k)\right]\nonumber\\
&&~~~~~+\epsilon_{ij}gf^{abc}\int_{p^+\gg k^+}A_{Fi}^b(k^++p^+,\v k+\v p)\alpha_{j}^{c\dagger}(k^+,\v k)\nonumber
\end{eqnarray}
\begin{eqnarray}
&&\frac{g}{2}\epsilon_{ij}f^{abc}\int A_i^b(p^+-k^+,\v p-\v k)A_j^c(k^+,\v k)\rightarrow\\
&&~~~g\epsilon_{ij}f^{abc}\int_{k\in\Omega_{DGLAP}} A_{Fi}^b(p^+-k^+,\v p-\v k)A_{Sj}^c(k^+,\v k)+g\epsilon_{ij}f^{abc}\int_{p^+\gg k^+} A_{Fi}^b(p^+-k^+,\v p-\v k)\alpha_{j}^c(k^+,\v k)\nonumber
\end{eqnarray}
Altogether
\begin{eqnarray}
B_{F}^a(p)&=&-\epsilon_{ij}\Bigg\{i\v p_iA^a_{Fj}(p^+,\v p) 
+gf^{abc}\int_{p^+\gg k^+;\v k^2\sim \v p^2}
\Big[A_{Fi}^b(p^+,\v k+\v p)\alpha_{j}^{c\dagger}(k^+,\v k)
\nonumber \\
&+&A_{Fi}^b(p^+,\v p-\v k)\alpha_{j}^c(k^+,\v k) \Big] +gf^{abc}\int_{k\in\Omega_{DGLAP}}\left[A_{Fi}^b(k^++p^+,\v k+\v p)A_{Sj}^{c\dagger}(k^+,\v k)
\right. \nonumber \\
&+&\left. A_{Fi}^{b\dagger}(k^+-p^+,-\v k-\v p)A_{Sj}^c(k^+,-\v k)
 +A_{Fi}^b(p^+-k^+,\v p-\v k)A_{Sj}^c(k^+,\v k )\right]\Bigg\}
\end{eqnarray}
The contribution of the soft fields can be conveniently encoded by defining the covariant derivative in the soft background  
\begin{equation}\label{Palpha}
P^{ab}_i\equiv \v p_i\delta^{ab}+igf^{abc}\int_{k^+\ll p^+; k^-\ll p^-}\left[\alpha_i^{\dagger c}(k^+,\v k)+\alpha_i^{ c}(k^+,-\v k)\right]\approx
 \v p_i\delta^{ab}+igf^{abc}\alpha_i^{c}(0)
\end{equation}
where $\alpha_i^c(0)$ is the soft field at the origin in coordinate space.
Then, neglecting consistently small momentum components throughout we can write
\begin{eqnarray}
\Pi_F^a(p)&=&-iP^{ab}_iA_{Fi}^b(p^+,\v p) +2g
f^{abc}\int_{k\in\Omega_{BFKL}}\frac{k^+}{p^+}A^{b\dagger}_{Sj}(k^++p^+,\v k-\v p) A^c_{Sj}(k^+,\v k) \nonumber\\
&+&g f^{abc}\int_{k\in\Omega_{DGLAP}}\left[A_{Si}^b(k^+,\v k)\frac{2k^+-p^+}{p^+}\left[A_{Fi}^{c\dagger}(k^+-p^+,\v k-\v p)+A_{Fi}^c(p^+-k^+,\v p-\v k)\right]\right.\nonumber\\
&&\left.
-A_{Si}^{b\dagger}(k^+,\v k)\frac{2k^++p^+}{p^+}A_{Fi}^c(p^++k^+,\v p+\v k)\right];\\ \nonumber\\
  B_F^a(p)&=&-i\epsilon_{ij}P_i^{ab}A^b_{Fj}(p)
 -\epsilon_{ij}gf^{abc}\int_{k\in\Omega_{DGLAP}}\left\{A_{Fi}^b(k^++p^+,\v k+\v p)A_{Sj}^{c\dagger}(k^+,\v k)
 \right. \nonumber \\
&& +\left.\left[A_{Fi}^{b\dagger}(k^+-p^+,-\v k-\v p) + A_{Fi}^b(p^+-k^+,\v p-\v k)\right]A_{Sj}^c(k^+,-\v k)\right\}\nonumber
\end{eqnarray}
As mentioned earlier, when substituted in the Hamiltonian ${\cal H}(p)$ (\ref{Hp}),
the first nonlinear term in $\Pi_F^a$ is obviously  the contribution of the standard eikonal interaction  between the fast and slow modes. The rest of the  terms, as we will see are responsible for the DGLAP collinear splittings, which clearly contribute in the frequency ordered approach.


Let us now examine the terms arising in ${\cal H}(p)$ when we substitute the above expressions for $\Pi_F$ and $B_F$.
First off, ${\cal H}(p)$ contains the  kinetic term for $A_F$ 
\begin{equation}
{\cal H}_0(p)={1\over 2}A^a_{Fi}(p^+,\v p)\left[\v P^2\delta_{ij}+[\v P_i,\v P_j]\right]^{ab}A^b_{Fj}(p^+,\v p)
\end{equation}
This involves the interaction of $A_F$ with the soft field $\alpha$ which here we are treating as a background. If this background is large it may significantly alter the dispersion relation of the fast mode. In the present pertrubative framework we are assuming that all fields are small, and therefore the change of dispersion relation itself is perturbative. 

At order $g$ we find two types of  interaction terms, responsible for the BFKL and DGLAP types of physics. The BFKL term is
\begin{equation}
{\cal H}_I^{BFKL}(p)=g\frac{1}{p^+}\left[A_{Fi}^{\dagger a}(p) \v P_i^{ab}\rho^b(p^+,-\v p)+A_{Fi}^{ a}(p) \v P_i^{ab}\rho^b(p^+,\v p)\right]
\end{equation}
The rest of the interaction Hamiltonian can be separated into several pieces.  
The most important is the term which is  naturally associated with the DGLAP evolution  
\begin{eqnarray}\label{dglap}
{\cal H}_I^{DGLAP}&=&-ig \int_{k\in\Omega_{DGLAP}}A^a_{Si}(k^+,\v k)f^{abc}\left[\delta_{ki}\delta_{jl}\left(\frac{2k^+}{p^+}-1\right)+\epsilon_{ki}\epsilon_{jl}\right]\v P^{bd}_jA^{\dagger d}_{Fl}(p^+,\v p)A^{\dagger c}_{Fk}(k^+-p^+,\v k-\v p)
\nonumber \\ &+&h.c.
\end{eqnarray}
This contains the vertex that creates two fast gluons from one slow mode, where the transverse momenta of the two created gluons are much higher than that of the slow parent one. This is of course the typical DGLAP splitting kinematics.

There are two more terms that differ from ${\cal H}_I^{DGLAP}$ only by the interchange of creation and annihilation operators
\begin{equation}\label{dglap1}
{\cal H}_I^{1}=-ig \int_{k\in\Omega_{DGLAP}}A^a_{Si}(k^+,\v k)f^{abc}\left[\delta_{ki}\delta_{jl}\left(\frac{2k^+}{p^+}-1\right)+\epsilon_{ki}\epsilon_{jl}\right]\v P^{bd}_jA^{\dagger d}_{Fl}(p^+,\v p)A^{ c}_{Fk}(p^+-k^+,\v p-\v k)+h.c.
\end{equation}
and
\begin{equation}\label{dglap2}
{\cal H}_I^{ 2}=-ig \int_{k\in\Omega_{DGLAP}}
A^{\dagger a}_{Si}(k^+,\v k)f^{abc}\left[\delta_{ki}\delta_{jl}\left(-\frac{2k^+}{p^+}-1\right)+\epsilon_{ki}\epsilon_{jl}\right]\v P^{bd}_jA^{\dagger d}_{Fl}(p^+,\v p)A^{ c}_{Fk}(p^++k^+,\v p+\v k)+h.c.
\end{equation}
These two terms, although present in the Hamiltonian, are irrelevant for finding the lowest order vacuum wave function of the fast modes, since they both vanish when acting on the zeroth order Fock vacuum of $A_F$.  For the purpose of calculating the evolution of the wave function we can thus ignore them. 
Were we interested in the excitations of $A_F$, these terms would certainly be important. 
In fact, as we will see soon they are also important in correctly determining the $S$ matrix consistent with our approximation for the evolution. In this section, however we are interested in the vacuum wave function of the fast modes, and will therefore only keep eq.\eqref{dglap}.


In addition to ${\cal H}(p)$, that is the interaction that arises from the square of the electric and magnetic fields of the fast modes, 
the Hamiltonian contains an interaction term that arises from the square of the electric (and magnetic) fields $\Pi_S^a(k) \Pi_S^a(k)$,
with low frequency $k^-$. It contains the eikonal interaction of the fast modes with the  soft fields $\alpha^a_i$:
\begin{equation}\label{dH}
\Delta {\cal H}(p)=
 -igf^{abc}A^{\dagger b}_{Fj}(p)A^c_{Fj}(p)\int_{k^+\ll p^+; k^-\ll p^-} {\v k_i\over k^+} \left[\alpha_{i}^{ a}(k^+,\v k)-\alpha_{i}^{\dagger a}(k^+,-\v k)\right]
\end{equation}
This term, just like ${\cal H}_I^{1,2}$  is irrelevant for determining the vacuum of the fast modes (in first order in perturbation theory), 
since it too annihilates the perturbative vacuum, and we disregard it in the following.

\subsection{The Interaction Hamiltonian and the Vacuum of  Fast Modes - more carefully}
\subsubsection{The Hamiltonian}
Collecting the basic expressions of the previous subsection we can write down the interaction Hamiltonian between fast and slow modes.  
From now on we will be more careful and will not allow double counting in the Hamiltonian. First off, to simplify notations we drop the subscripts $F$ and $S$ on the fields.
Instead we will simply stick to the convention that the momentum of mode with highest frequency that appears in the interaction Hamiltonian is denoted by $p$.
Also, as discussed above, for the purpose of finding the ground state we only need the terms in  the interaction Hamiltonian that create the gluon with the highest frequency.

The kinetic term for the fast modes is given by
\begin{equation}
H_0=\int_p {\cal H}_0(p)\,; \qquad\qquad
{\cal H}_0(p)=\frac12A^a_{i}(p^+,\v p)\left[\v P^2\delta_{ij}+[\v P_i,\v P_j]\right]^{ab}A^b_{j}(p^+,\v p).
\end{equation}
The interaction Hamiltonian is
\begin{eqnarray}\label{inter}
H_I=\int_p {\cal H}_I(p); \qquad  &&{\cal H}_I(p)=
-ig \int_{max({k^-,(k-p)^-)}<p^-}A^a_{i}(k^+,\v k)f^{abc}\,\times\nonumber \\
&&\times\left\{\left[\delta_{ki}\delta_{jl}\left(\frac{2k^+}{p^+}-1\right)+\epsilon_{ki}\epsilon_{jl}\right]\v P^{bd}_jA^{\dagger d}_{l}(p^+,\v p)A^{\dagger c}_{k}(k^+-p^+,\v k-\v p)
\right. \\
&&+\left.\left[\delta_{ki}\delta_{jl}\left(\frac{2k^+}{k^+-p^+}-1\right)+\epsilon_{ki}\epsilon_{jl}\right](\v K-\v P)^{bd}_jA^{\dagger d}_{l}(k^+-p^+,\v k-\v p)A^{\dagger c}_{k}(p^+,\v p)\right\}
+h.c. \nonumber
\end{eqnarray}
As discussed above this interaction Hamiltonian contains the eikonal interaction (in the first term) in the part of the phase space where $p^+\ll k^+$, and the DGLAP splittings (in both terms) for $|\v p|\gg |\v k|$. Since in \eqref{inter} we do not explicitly separate between the BFKL and DGLAP interactions, the whole phase space of the interaction is included without double counting.

Our next goal is to find the wave function of the high frequency modes which is the ground state of this Hamiltonian. At this point, as per discussion above we 
drop the soft fields from consideration, and replace the covariant derivatives in the soft background by simple derivatives. In addition the above expression can be simplified somewhat by recognizing that the second term is only important for DGLAP splittings, where $|\v p|\gg |\v k|$. We therefore write
\begin{eqnarray}\label{inter1}
{\cal H}_I(p)&\approx&-ig \int_{max({k^-,(k-p)^-)}<p^-}A^a_{i}(k^+,\v k)f^{abc}\,\times\nonumber \\
&\times&\left\{\left[\delta_{ki}\delta_{jl}\left(\frac{2k^+}{p^+}-1\right)+\epsilon_{ki}\epsilon_{jl}\right]\v p_jA^{\dagger b}_{l}(p^+,\v p)A^{\dagger c}_{k}(k^+-p^+,\v k-\v p)
\right.\nonumber \\
&-&\left.\left[\delta_{ki}\delta_{jl}\left(\frac{2k^+}{k^+-p^+}-1\right)+\epsilon_{ki}\epsilon_{jl}\right]\v p_jA^{\dagger b}_{l}(k^+-p^+,\v k-\v p)A^{\dagger c}_{k}(p^+,\v p)\right\}
+h.c.
\end{eqnarray}
and will use this interaction Hamiltonian in our calculations.

\subsubsection{The Light Cone Wave Function}

To find the vacuum wave function of the fast modes we use the standard perturbative procedure. 
To leading  order in perturbation theory the operator that diagonalizes the Hamiltonian ${\cal H}_I(p)$ of the fast mode with momentum 
$p$ can be written as
\begin{equation}
\Omega_p=1+iG(p^+,\v p)\approx e^{iG(p^+,\v p)}; \qquad\qquad G(p^+,\v p)= {\cal H}_I(p) D
\end{equation}
where $G(p^+,\v p)$ is the interaction Hamiltonian for the mode $p$ multiplied by the energy denominator $D$.
In calculating the latter  we can  make the following  simplification
\begin{equation}\label{den}
D^{-1}\equiv k^--p^--(k-p)^-=\frac{  \v k^2}{2k^+}-\frac{  \v p^2}{2p^+}-\frac{(  \v k- \v p)^2}{2(k^+-p^+)}\approx-\frac{  \v p^2k^+}{2p^+(k^+-p^+)}
\end{equation}
This expression is obviously valid in the BFKL regime, where we can neglect $k^-$ and $(k-p)^-$ relative to $p^-$. In the DGLAP regime one gluon splits into two gluons of higher frequency, and the frequency of those two, $p^-$ and $(k-p)^-$ may be comparable. Hence in this regime we can neglect $k^-$, and assume $\v p^2\gg \v k^2$, which again leads to \eqref{den}.
Thus the approximation \eqref{den} encompasses both, the BFKL regime ($k^+\gg p^+,  \   \v p^2\sim   \v k^2$) and the DGLAP regime ($k^+\sim p^+,\   \v k^2\ll  \v p^2$).
Using \eqref{den} we can represent the diagonalizing operator in the form
 \begin{equation}
    \begin{split}
    G(p^+, \v p)&=g \int_{k^-<p^-; \ (k-p)^-<p^-}
    A^a_{i}(k^+,\v k)\frac{2p^+(k^+-p^+)}{k^+}\, \\
&\times\left\{\left[\delta_{ki}\delta_{jl}\left(\frac{2k^+}{p^+}-1\right)+\epsilon_{ki}\epsilon_{jl}\right]\frac{  \v p_j}{\v  p^2} A^{\dagger }_{l}(p^+,\v p)T^a A^{\dagger }_{k}(k^+-p^+,\v k-\v p)
\right. \\
&-\left.\left[\delta_{ki}\delta_{jl}\left(\frac{2k^+}{k^+-p^+}-1\right)+\epsilon_{ki}\epsilon_{jl}\right]\frac{  \v p_j}{ \v p^2} A^{\dagger }_{l}(k^+-p^+,\v k-\v p)T^a A^{\dagger}_{k}(p^+,\v p)\right\} 
+h.c.
\end{split}
\end{equation}
A representation that we will find convenient is
\begin{equation}\label{ca}
\begin{split}
G(p^+, \v p)= A^{\dagger }_{i}(p^+,  \v p)C_i(p^+,  \v p)+  A_{i}(p^+,  \v p)C_i^\dagger(p^+,  \v p)
\end{split}
\end{equation}
with  
\begin{equation}\label{c}
\begin{split}
C_i^{a}(p^+,   \v p)
=&g \int_{p^->k^-, p^->(k-p)^-} 
  \frac{2p^+(k^+-p^+)}{k^+}\, \\
&\times\left\{\left[\delta_{kl}\delta_{ji}\left(\frac{2k^+}{p^+}-1\right)+\epsilon_{lk}\epsilon_{ji}\right]+\left[\delta_{ki}\delta_{jl}\left(\frac{2k^+}{k^+-p^+}-1\right)+\epsilon_{ik}\epsilon_{jl}\right]\right\} \\
&\times \frac{ \v p_j}{ \v p^2} A^{\dagger}_{l}(k^+-p^+, \v k- \v p)T^a  A_{k}(k^+, \v k)
\end{split}
\end{equation}

The operator $C_i^a$ reduces to the "classical field" produced by the slow modes  in the BFKL regime, $p^+\ll k^+$. In  the DGLAP regime, for $| \v p|\gg | \v k|$ and $k^+\sim p^+$ it contains one slow and one fast mode but we still find this form useful in calculations.
As a shorthand, it is convenient to introduce
\begin{equation}\label{F}
\begin{split}
F_{lk}^i(k,p)&=\frac{2p^+(k^+-p^+)}{k^+}\left\{\left[\delta_{kl}\delta_{ji}\left(\frac{2k^+}{p^+}-1\right)+\epsilon_{lk}\epsilon_{ji}\right]+\left[\delta_{ki}\delta_{jl}\left(\frac{2k^+}{k^+-p^+}-1\right)+\epsilon_{ik}\epsilon_{jl}\right]\right\} \frac{ \v p_j}{\v  p^2} \\
&=\frac{4p^+(k^+-p^+)}{k^+}\left\{\delta_{kl}\delta_{ji}\frac{k^+}{p^+}+\delta_{ki}\delta_{jl}\frac{k^+}{k^+-p^+}-\delta_{kj}\delta_{il}\right\} \frac{\v p_j}{\v p^2} 
\end{split}
\end{equation}
where one simply used the identity, $\epsilon_{ij}\epsilon_{kl}=\delta_{ik}\delta_{jl}-\delta_{il}\delta_{jk}$, so that
\begin{equation}
\begin{split}
 C_i^{a}(p^+, \v p)&=g\int_{k^-<p^-;\ \ (k-p)^-<p^-}
  F^i_{lk}(k,p) A^{\dagger}_{l}(k^+-p^+, \v k-\v  p)T^a  A_{k}(k^+,  \v k)
\end{split}
\end{equation}
\begin{equation}
\begin{split}
 C_i^{a\dagger}(p^+, \v p)&=g\int_{k^-<p^-;\ \ (k+p)^-<p^-}
 F^i_{kl}(k+p,p) A^{\dagger}_{l}(k^++p^+,  \v k+ \v p)T^a  A_{k}(k^+,\v  k)
\end{split}
\end{equation}

Having diagonalized the Hamiltonian for the single fast mode $p$,
it is now straightforward to write down the wave function of the entire projectile system evolved from the initial frequency 
 $E_0$ to the frequency $E$: 
\begin{equation}\label{psiel1}
 |\Psi_P\rangle_E={\cal P}\exp\left\{i\int_{E_0}^{E}d p^- {\cal G}(p^-)\right\}|\Psi_P\rangle_{E_0}
 \end{equation}
where ${\cal P}$ denotes path ordering and 
\begin{eqnarray}
{\cal G}(p^-)&\equiv& \int_p 
\delta(p^--\frac{\v p^2}{2p^+})G(p^+,\v p^2); 
 \end{eqnarray}
 with $G(p^+,\v p)$ given by \eqref{ca}, \eqref{c}.
 Since the exponential operator in \eq{psiel1} is unitary, the wave function is properly normalized.

The wave function \eqref{psiel1} solves the problem we have posed to ourselves to find the LCWF of fast modes in BO approximation.

\subsection{ BFKL vs DGLAP}
The evolved wave function of the system \eqref{psiel1} is the basis for calculations of the frequency evolution of various observables. These calculations will be presented in the subsequent papers in the series. Most of these calculations are performed directly using the form \eqref{ca}. However sometimes it is useful to separate different contributions explicitly into those due to the BFKL and DGLAP type processes. In the previous section we have discussed rather loosely how the two parts of the phase space are reflected in the interactions between the slow and fast modes. 
 So far we have not  split BFKL and DGLAP regions of the space space carefully, since in the DGLAP term we loosely are assuming that the longitudinal momenta of all interacting gluons are of the same order. However technically we are integrating over all values of transverse momenta, and therefore there is a potential (and actual) double counting here since the BFKL and DGLAP kinematic regions the way we have been treating them so far, overlap. 
 
 For the purpose of calculation however we need to carefully delineate the separation between the two parts of the phase space in order to avoid any double counting. In this subsection we do exactly that: we will define precisely what we mean by the BFKL region and the DGLAP region.

 Consider again the triple gluon interaction vertex between the slow and fast modes. As before we denote by $(\v p,\ p^+)$ the momentum of the mode with the highest frequency, and also specify that  $(\v k,\ k^+)$ are the transverse and longitudinal momenta of the field mode with the lowest frequency,  so that $E\gtrsim p^->(k-p)^->k^-$. Clearly the field with momentum $k$ is the "slow mode" while that with momentum $p$ is the "fast mode". The third field entering the vertex may be in principle either slow (closer to $k^-$) or fast (closer to $p^-$), but for now we are not concerned with it. Our goal is to separate in an unambiguous way the emission vertex of $p$ from $k$ into the DGLAP and BFKL parts. In the BFKL part of the vertex we can then use the standard eikonal approximation, i.e. will neglect $p^+$ relative to $k^+$, while in the DGLAP part we can use the collinear approximation, i.e. neglect $\v k$ relative to $\v p$.
 
It is clear that the boundary between the BFKL and DGLAP regions should lay somewhere in the Doubly Logarithmic (DLA)  region which is where both, eikonal and collinear approximations are valid. 

From the eikonal approximation point of view it is natural to consider the boundary to be along a line   $p^+=\lambda k^+$, where $\lambda$ is a small number $\lambda<1$, but which  parametrically may be of order unity,
with the BFKL region restricted to $p^+<\lambda k^+$. 

From the collinear approximation (DGLAP) point of view, the natural separator is a line associated with the ordering in transverse momenta,  $\v k^2=\delta \v p^2$, $\delta<1$, with the DGLAP region restricted to  $\v k^2< \delta \v p^2$.  The two lines $p^+=\lambda k^+$ and $\v k^2=\delta \v p^2$ are distinct unless we choose $\lambda$ and $\delta$ to be frequency dependent in the following simple way
\beq
\lambda=\delta= \sqrt{\frac{k^-}{p^-}}
\eeq  
For this choice the unique separation line is
\beq
\frac{k^+}{p^+}=\frac{\v p^2}{\v k^2}
\eeq
The phase space is then separated into two complementary regions
\beq
{\rm BFKL:} ~~ k^+ > p^+ \sqrt{p^-\over k^-} \  \ \ \Bigg[{\rm or}\ \ \v p^2<\v k^2\sqrt{\frac{p^-}{k^-}}\Bigg];~~~~~~~~~~~ {\rm DGLAP: }~ ~\v p^2>\v k^2\sqrt{\frac{p^-}{k^-}} \ \ \ \ \Bigg[{\rm or} \ \ \ k^+ < p^+ \sqrt{p^-\over k^-} \Bigg]
\eeq  
Since by definition $p^->k^-$, this separation indeed achieves its goal,  in the sense that for all momenta we have $\lambda<1$ and $\delta<1$.


With this separation we can decompose the interaction Hamiltonian into two terms.
\beq
H_I=H_I^{BFKL}+H_I^{DGLAP}
\eeq
 where the BFKL Hamiltonian is
\begin{equation}
H_I^{BFKL}=g\int_p\frac{1}{p^+}\left[A_{i}^{\dagger a}(p) \v p_i\rho^a(p^+,-\v p)+A_{i}^{ a}(p) \v p_i\rho^{*a}(p^+,-\v p)\right]
\end{equation}
with
 \begin{equation}\label{rho+}
\rho^a(p^+,\v p)= -2i
f^{abc}\int_{k^+> \sqrt{\frac{p^-}{k^-}}p^+,\, max(k^-, (k+p)^-)<p^- } \,k^+\,A^{b\dagger}_{j}(k^+,\v k+\v p) A^c_{j}(k^+,\v k)\ ,
\end{equation}
and the DGLAP  Hamiltonian is
\begin{eqnarray}
H_I^{DGLAP}&=&-ig \int_{\v p^2>\sqrt{\frac{p^-}{k^-}} \v k^2;\, p^+<k^+/2;\, k^-<p^-}A^a_{i}(k^+,\v k)f^{abc}\,\times\nonumber \\
&\times&\left\{\left[\delta_{ki}\delta_{jl}\left(\frac{2k^+}{p^+}-1\right)+\epsilon_{ki}\epsilon_{jl}\right]\v p_j A^{\dagger b}_{l}(p^+,\v p)A^{\dagger c}_{k}(k^+-p^+,-\v p)
\right.\nonumber \\
&-&\left.\left[\delta_{ki}\delta_{jl}\left(\frac{2k^+}{k^+-p^+}-1\right)+\epsilon_{ki}\epsilon_{jl}\right]\v p_j A^{\dagger b}_{l}(k^+-p^+,-\v p)A^{\dagger c}_{k}(p^+,\v p)\right\}
+h.c.
\end{eqnarray}


Finally, the evolved wave function can be written as
\begin{equation}\label{psiel}
 |\Psi_P\rangle_E={\cal P}\exp\left\{i\int_{E_0}^{E}d p^- \left[{\cal G}^{BFKL}(p^-)+{\cal G}^{DGLAP}(p^-)\right]\right\}|\Psi_P\rangle_{E_0}
 \end{equation}
where
\begin{eqnarray}
{\cal G}^{BFKL}(p^-)&\equiv& \int_p\ \delta(p^--\frac{\v p^2}{2p^+})G^{BFKL}(p^+,\v p); \nonumber\\
 {\cal  G}^{DGLAP}(p^-)&\equiv& \int_p \ \delta(p^--\frac{\v p^2}{2p^+})G^{DGLAP}(p^+,\v p);
 \end{eqnarray}
with
\begin{eqnarray}\label{gbfkl}
G^{BFKL}(p^+,\v p)&=&2g \left[A_{i}^{\dagger a}(p) \frac{\v p_i}{\v p^2}\rho^a(p^+,-\v p)+A_{i}^{ a}(p) \frac{\v p_i}{\v p^2}\rho^{\dagger a}(p^+,-\v p)\right]
\end{eqnarray}
and
\begin{eqnarray}\label{gdglap}
 G^{DGLAP}(p^+,\v p)&=&-ig \int_{p^2>\sqrt{\frac{p^-}{k^-}}k^2;\, p^+<k^+/2;\, k^-<p^-} 
 f^{abc}A^a_{i}(k^+,\v k)\frac{2p^+(k^+-p^+)}{k^+}
\\
 &\times&\Bigg\{ \left[\delta_{ki}\delta_{jl}\left(\frac{2k^+}{p^+}-1\right)+\epsilon_{ki}\epsilon_{jl}\right]\frac{\v p_j}{\v p^2}A^{\dagger b}_{l}(p^+,\v p)A^{\dagger c}_{k}(k^+-p^+,-\v p)\nonumber\\
 &+&\left[\delta_{li}\delta_{jk}\left(\frac{2k^+}{k^+-p^+}-1\right)+\epsilon_{li}\epsilon_{jk}\right]\frac{\v p_j}{\v p^2}A^{\dagger b}_{l}(p^+,\v p)A^{\dagger c}_{k}(k^+-p^+,-\v p)\Bigg\}+h.c.\nonumber
 \end{eqnarray}
With this separation there is no overlap between the BFKL and DGLAP region on one hand, and all the phase space is covered on the other.

\section{The scattering matrix}

Our eventual goal of course is to calculate the evolution of physical observables. The simplest such observable, which is the object of the standard BFKL evolution is the total cross section, or the forward scattering amplitude of hadronic scattering.  We therefore have to understand what is the form of the $S$-matrix operator appropriate to our approximation. 
Such a consistency between the approximation to the evolved wave function and the $S$-matrix must exist. For example, in the standard BFKL and JIMWLK treatment the scattering matrix is taken to be eikonal, which is consistent with the approximation that kept only eikonal vertices in the Hamiltonian used to evolve the wave function. Here we are not limiting ourselves to the eikonal vertices, but are also keeping the DGLAP splittings as appropriate to frequency ordering, and we should therefore consistently derive  the $S$-matrix in the same approximation.

In general the consistency stems from the fact that the separation between the projectile and target degrees of freedom is done in the same way as the separation between the slow and fast modes in the projectile wave function.

 The $S$-matrix operator has been formally defined in \eq{S}.  Its forward matrix element is given by
 \begin{equation}
 S= \langle \Psi_{in} |\hat S|\Psi_{in}\rangle\,=\,\langle\Psi_P|\langle \Psi_T|\,\hat S\,|\Psi_T\rangle |\Psi_P\rangle
 \end{equation}
Formally averaging  over the target wave function we obtain the $S$-matrix operator acting on the projectile Hilbert space,
 \begin{equation}
 \hat S_P=\langle \Psi_T|\hat S|\Psi_T\rangle\,.
 \end{equation}
 This is what we ordinarily do in the JIMWLK framework with the assumption that the target fields are large, which allows us to substitute classical fields for the quantum target field operators in  $\hat S_P$. Our goal here is also to derive the form of the  $S$-matrix operator acting on the projectile Hilbert space $\hat S_P$.
 
Conceptually, the form of the $S$ matrix should be similar to the form of the LCWF. As we discussed above, the $S$-matrix is given by the infinite time limit of the time evolution operator of the projectile-target system \eqref{S}, while the LCWF is given by the action of the diagonalizing operator $\Omega$, which is also the infinite time limit of the time evolution operator, but it is time of the fast-slow mode system in the projectile \eqref{usf},\eqref{omega}. Since the separation between the projectile and target on one hand, and between slow and fast modes on the other both are done on the basis of frequency, one expects that the two evolution operators are very similar.

There are a couple of important points here that prevent us from writing down $\hat S$ immediately once the form of $\Omega$ is know. First, in the calculation of the evolution we are acting with the operator $\Omega$ on the vacuum of the fast modes, and can therefore neglect the operators of the type ${\cal H}_I^{1,2}$ (\ref{dglap1},\ref{dglap2}) which annihilate the vacuum. On the other hand in calculating the $S$-matrix the Hamiltonian acts on an unspecified, but not empty 
target state.  Thus such terms cannot be neglected. 

Second, and more critical is that since we are working in the projectile light cone frame, the target degrees of freedom do not appear as on shell gluon modes at all. Instead they appear as the infamous zero modes on the light front and their dynamics is not contained in the light front projectile Hamiltonian. Nevertheless the light front Hamiltonian does contain their interaction with the on shell projectile gluons, and this is what determines the operator form of the $S$-matrix. 
 
In this paper we will not consider dense targets. This is left for future work.  Our goal now is to derive the explicit form of the
  $S$-matrix operator for a dilute target. That is we assume that the target fields are weak and the $S$-matrix can be expanded to second order in target fields. We will now derive  as an operator in the projectile Hilbert space. To derive $\hat S_P$ we will go back to the projectile light front Hamiltonian and perform a more careful quantization keeping the zero modes which were neglected in Section 2. Those were unimportant as long as we were interested in dynamical projectile gluons only, but we now need to keep them alive in order to determine  the interaction between the projectile gluons and the target degrees of freedom. 
  . 
  
  \subsection{The $S$-matrix in the dilute target limit}

In (\ref{LYM}) the kinetic term of the transverse field is given by $\partial^-A^i\partial^+A^i$. This indicates that the mode of the transverse field with zero longitudinal momentum $p^+$ is  non-dynamical, reflecting the fact  that after fixing $A^+=0$ gauge there still remains a residual gauge invariance under $x^-$ independent gauge transformations. As noted above, this $p^+=0$  mode should be understood as representing the target field, and therefore indeed it's dynamics should be contained in the target Hamiltonian. The light cone Hamiltonian of the projectile determines only the interaction of the target fields with the projectile gluons.
Without further ado will simply assume that the dynamics of the target is prescribed. In practical terms we can therefore treat the zero modes as an external field which is distributed with some probability distribution determined by the target wave function. In the following we denote the modes of $A^\mu$ with nonzero momentum by $\tilde A^\mu$, while zero modes by $ \gamma^\mu$,
 Separating the zero modes,  the nonzero longitudinal momentum modes of the field strength tensor read
 \begin{eqnarray}
\tilde F_a^{+i}&=&\partial^+ \tilde A_a^i \nonumber \\
\tilde F_a^{-i}&=&\Big(\partial^- \tilde A_a^i -\partial^i \tilde A_a^- -gf^{abc}\tilde A_b^-\tilde A_c^i \Big)-gf^{abc}\tilde A_b^-  \gamma_c^i-gf^{abc} \gamma_b^-\tilde A_c^i \nonumber \\
\tilde F_a^{ij}&=&\partial^i \tilde A_a^j-\partial^j \tilde A_a^i-gf^{abc}\tilde A_b^i \gamma_c^j-gf^{abc} \gamma_b^i\tilde A_c^j-gf^{abc}\tilde A_b^i\tilde A_c^j.
\end{eqnarray}
As mentioned above, we assume that the target fields are small, and therefore we only need to find the interaction to first order in $\gamma$. To this order  the Lagrangian density (\ref{LYM}) explicitly becomes
 \begin{eqnarray}
 &&{\cal L}_{YM}=\frac{1}{2}(\partial^+\tilde A_a^-)^2+\partial^+\tilde A_a^i\Big(\partial^- \tilde A_a^i -\partial^i \tilde A_a^- -gf^{abc}\tilde A_b^-\tilde A_c^i \Big)-\frac{1}{4}\Big(\partial^i \tilde A_a^j-\partial^j \tilde A_a^i-gf^{abc}\tilde A_b^i\tilde A_c^j\Big)^2\\
 &&-\partial^+ \tilde A_a^i \Big(gf^{abc}\tilde A_b^- \gamma_c^i+gf^{abc} \gamma_b^-\tilde A_c^i\Big)+gf^{abc} \gamma_b^i\tilde A_c^j \Big(\partial^i \tilde A_a^j-\partial^j \tilde A_a^i-gf^{ade}\tilde A_d^i\tilde A_e^j\Big) +\frac{1}{2}gf^{abc}\tilde A_b^i\tilde A_c^j
(\partial^i \gamma_a^j-\partial^j  \gamma_a^i)\nonumber
\end{eqnarray}
 The first line  defines the dynamics of $\tilde A$ which has been explored in the previous section. The second line represents its
 interaction with the zero mode.
 
The non-dynamical field component $\tilde A^-$  is eliminated by using the constraint equation which follows from differentiating the Lagrangian with respect to $\tilde A^-$:
 \begin{equation}\label{-con}
 -(\partial^+)^2\tilde A_a^-+\partial^i\partial^+\tilde A_a^i-gf^{abc}(\tilde A_b^i+ \gamma_b^i)\partial^+\tilde A_c^i=0
 \end{equation}
 Note that on-shell eq.\eqref{-con} is identical to the nonzero mode part of eq.\eqref{+con} (here the on-shell  means on solutions of the equations of motion for the transverse fields, which determine the dispersion relation for $\tilde A^i$), so that this equation is redundant. 
 The zero mode $\gamma^-$ is determined directly by eq.\eqref{+con} by taking its zero longitudinal momentum component, as
 \begin{equation}
 \gamma^-_a=\frac{\partial_i}{\partial^2}\partial^-\gamma^i_a+O(g)
 \end{equation}
 We will use this relation later on to express the interaction with the target entirely in terms of the fields $\gamma^i_a$.

 The Hamiltonian for the nonzero modes, i.e. for the projectile gluons, is given by the Legendre transform of the Lagrangian. 
 In the absence of zero modes, the result of the Legendre transform is the usual light front Hamiltonian that we have used in the previous sections. Here we are only interested in the interaction term linear in the zero mode. Keeping only terms that arise from the three gluon coupling,  we find
 \begin{eqnarray}
 H_{PT}&=&
gf^{abc} \int d^3x \left\{ \partial^+ \tilde A_a^i \Big(\tilde A_b^- \gamma_c^i+ \gamma_b^-\tilde A_c^i\Big)- \gamma_b^i\tilde A_c^j 
\Big(\partial^i \tilde A_a^j-\partial^j \tilde A_a^i\Big) -\frac{1}{2}\tilde A_b^i\tilde A_c^j
(\partial^i \gamma_a^j-\partial^j  \gamma_a^i)\right\}\nonumber\\
&=&gf^{abc}\int d^3x\left\{- \gamma_a^i\tilde A_b^i\partial^j\tilde A_c^j-\partial^i \gamma_a^j\tilde A_b^i\tilde A_c^j
- \gamma_b^i\tilde A_c^j \Big(\partial^i \tilde A_a^j-\partial^j \tilde A_a^i\Big)+\left[\partial^-\frac{\partial^i}{\partial^2}\gamma^i_a\right]
\tilde A^j_b\partial^+\tilde A^j_c\right\}\nonumber\\
&=&gf^{abc} \int d^3x\left\{\left[\partial^-\frac{\partial^i}{\partial^2} \gamma^i_a\right]\tilde A^j_b\partial^+\tilde A^j_c- \gamma^i_a\tilde A^j_b\partial^i\tilde A^j_c+2[\partial^j \gamma_a^i]\tilde A^i_b\tilde A^j_c\right\}
\end{eqnarray}
 Above we have used, that to this order
 \begin{equation}
 \frac{1}{2}(\partial^+\tilde A_a^-)^2-\partial^i\tilde A_a^i \partial^+\tilde A_a^- =-\frac{1}{2}(\partial^i\tilde A_a^i)^2
 \end{equation}
The terms we keep involve two projectile gluons and one target field. This naturally describes scattering of a projectile gluon via a single scattering on the target. Notice that, since we treat $\gamma$ as a field with externally prescribed dynamics, it is in general time $(x^+)$ dependent. The interaction of projectile gluons with the zero mode, makes the Hamiltonian $H_{PT}$  time dependent as well.

 We next write the Hamiltonian in  the Fourier space.
 For the nonzero modes we use (\ref{Ak}).
 For the zero mode we write
 \begin{equation}
   \gamma^i_a(x^+,  \v x)=\int \frac{dq^-d^2  q}{(2\pi)^3}\left[ \gamma^i_a(q^-,  \v q)e^{-iq\cdot  x}+\gamma^{* i}_a(q^-,  \v q)e^{iq\cdot  x}\right]
  \end{equation}
  with $q\cdot  x\equiv q^-x^+-  \v q \cdot  \v x$.
 
  The interaction Hamiltonian is 
 \begin{eqnarray}\label{hs}
 H_{PT}(x^+)&=& 
 igf^{abc}\int \frac{d^3k}{(2\pi)^3 }\frac{d q^-}{2\pi}\frac{d^2  \v q}{(2\pi)^2}e^{-i D^{-1} x^+} \gamma^i_a(q)\tilde A^{\dagger j}_b( k+  q)\tilde A^k_c(k) \nonumber \\
&\times&\left[\delta^{jk}\left(\frac{2k^+q^-}{  \v q^2}  \v q^i-(2  \v k^i+  \v q^i)\right)+2\epsilon^{il}  \v q^l\epsilon^{jk} \right]+h.c
 \end{eqnarray}
 where\footnote{A proper $i\epsilon$ prescription is implied here.}
 \begin{equation}\label{denom}
 D=\frac{1}{q^-+\frac{  \v k^2}{2k^+}-\frac{(  \v k+  \v q)^2}{2k^+}}
 \end{equation}
 
 The $S$-matrix operator is given by the path ordered exponential over the light cone time $x^+$, 
 for which (in the leading perturbative approximation
 \footnote{Beyond the leading order we would have to be careful about the path ordering.}) we need to integrate the interaction picture Hamiltonian $H^I_{PT}$, \eq{S}.  The time dependence is explicit in (\ref{hs}).
 The result of the integration brings in the energy denominator $D$, \eqref{denom}.

 This can be further simplified.  
 First we note that by assumption $q^-\gg k^-$, and so the second term in the denominator can be dropped. As for the frequency of the scattered particle, $(k+q)^-$, it can be either much smaller than the frequency of the target field, in which case it can be neglected, or of the same order. The latter situation can only be achieved for $  \v q^2\gg  \v k^2$. We can then approximate the energy denominator in all relevant regions by
 \begin{equation}
 D\approx\frac{2k^+}{2q^-k^+-  \v q^2}
 \end{equation}
 By the same token we can neglect the $\v k^i$ term in \eqref{hs} relative to $\v q^i$, since it is not leading neither in the BFKL, nor in the DGLAP regime. Finally we can write 
the $S$-matrix operator as
\begin{equation}\label{Sp}
\hat S_P=\langle\Psi_T|\exp\left\{i\int_q
\tilde G(q)\right\}|\Psi_T\rangle
\end{equation}
 with
 \begin{equation}
 \tilde G(q)=gf^{abc}\int_{k:\ k^-<q^-} 
 \gamma^a_i(q)f^i_{jk}(k,q)\tilde A^{\dagger j}_b(k^+,  \v k+ \v q)\tilde A^k_c(k^+,\v  k)+h.c.
 \end{equation}
 where
 \begin{equation}\label{fijk}
 f^i_{jk}(k,q)=2k^+\frac{  \v q^i}{\v  q^2}\delta^{jk} +4k^+\epsilon^{il}\frac{  \v q^l}{  \v q^2}\epsilon^{jk} \frac{1}{\frac{2q^-k^+}{  \v q^2}-1}
 \end{equation}
 
 The averaging indicated in \eqref{Sp} is the averaging over the target wave function, which has to be specified "externally" in the sense that it is unrelated to the discussion of the projectile wave function given in this paper. For our purposes any quantum properties of this averaging procedure are unimportant and we can simply think of it as averaging over some specified statistical ensemble of the target fields $\gamma$.
 
 For weak target fields we can expand \eqref{Sp} in powers of $\gamma$. At first order, this expansion does not contribute to the forward scattering, as the operator $\tilde G$ is a color octet in the projectile and target degrees of freedom separately.  In the weak target field limit therefore, the $S$-matrix can be written as (assuming color and rotational invariance of the target and/or projectile wave function\footnote{By assuming rotational invariance we restrict ourselves to unpolarized scattering. One can of course relax this assumption and include polarized scattering into consideration, but this is not our purpose here.})
 \begin{equation}\label{Sga}
 S=1\,-\,{1\over 2(N_c^2-1)}\,\int_q 
 \langle\Psi_T| \gamma^{\dagger b}_j(q)\gamma^b_j(q)|\Psi_T\rangle\,
 O(q)
 \end{equation}
 where $\langle\Psi_T|\gamma^{\dagger b}_j(q)\gamma^b_j(q)|\Psi_T\rangle$ denotes the average over the target wave function (ensemble of fields), and  $O$ is the matrix element in the projectile Hilbert space
 \begin{eqnarray}\label{scat}
&&O(q)=\langle \Psi_P |{\cal C}_i^{a} (q^-,\v q){\cal C}^{a\dagger}_i(q^-,\v q)|\Psi_P\rangle_E;\\ \nonumber \\
&&{\cal C}_i^{a\dagger}(q^-,\v q)=g\int_{k:k^-<q^-}
f^i_{lk}(k,q) A^{\dagger}_{l}(k^+,  \v k+ \v q)T^a  A_{k}(k^+,\v  k)
\end{eqnarray}
 Our formalism  of the projectile-target separation in the frequency assumes that  the frequency distribution of the target fields cuts off the frequencies from below  at $E$, i.e. $q^->E$, where $E$ has been introduced as the maximal frequency of gluon modes included in the projectile wave function. 
 It is in fact convenient to think of $\langle\Psi_T|\gamma^{\dagger b}_j(q)\gamma^b_j(q)|\Psi_T\rangle$ as being sharply peaked around $q^-\simeq E$, although this assumption is not strictly necessary.
 
 We recognize in the expressions \eqref{Sga}, \eqref{scat} and \eqref{fijk} familiar limiting cases. For example 
the BFKL (eikonal) approximation for the scattering matrix is recovered if we assume that the frequency of the target field is very high, i.e.   $q^-\gg \frac{ \v q^2}{2k^+}$. Then we have
\begin{equation}
f^i_{lk}(k,q)\rightarrow 2k^+\frac{  \v q_i}{\v  q^2}\delta_{kl}; \ \ \ \ \ \quad {\cal C}_i^{a\dagger}(q^-,q)\rightarrow \frac{  \v q_i}{\v  q^2}\rho^a(\bar q^+, \v q); \ \ \ \ \bar q^+\equiv\frac{\v q^2}{2q^-}
\end{equation}
In the opposite limit of very high transverse momentum exchange, the scattering matrix is not eikonal and corresponds to hard scattering from the target.
In this limit the second term in \eqref{scat} dominates
\begin{equation}\label{nbfkl}
f^i_{jk}(k,q)\rightarrow 4k^+\epsilon^{il}\frac{  \v q^l}{\v  q^2}\epsilon^{jk} \frac{1}{\frac{k^+}{  \bar q^+}-1}\,=\,
 4k^+ \frac{\v q^k\delta^{ij}-\v q^j\delta^{ik}}{{2q^-k^+}-{  \v q^2}}.
 \end{equation}
We will see in the next subsection that this limit of $f$ indeed corresponds to DIS of a probe that directly couples to gluons.

\subsection{Single and double particle scattering contributions}
 
The scattering amplitude  \eqref{scat} can be separated into two physically rather distinct contributions. One can see this by considering the normal ordered form of \eqref{scat}.
\begin{equation}
O(q)=O_1(q)+O_2(q)
\end{equation}
with 
\begin{equation}\label{1ps}
O_1(q)=g^2N_c \int_{k:k^-<q^-} 
\frac{1}{2k^+}f^i_{lk}(k,q)f^i_{lm}(k,q)
\langle \Psi_P | A^{a\dagger}_k(k^+, \v k)A^a_{m}(k^+,  \v k)|\Psi_P\rangle_E
\end{equation}
\begin{eqnarray}\label{2ps}
O_2(q)
&=&g^2\int_{\{k,l:\ (k-q)^-<q^-;k^-<q^-;\ (l-q)^-<q^-; l^-<q^-\}}
 f^i_{lk}(k,q)  f^i_{nm}(l,q)\nonumber\\ \nonumber \\
&\times& \langle \Psi_P | :  A^{\dagger}_{l}(k^+,  \v k- \v q)T^a  A_{k}(k^+, \v k) A^{\dagger}_{m}(l^+, \v l+ \v q)T^a  A_{n}(l^+,\v  l) :|\Psi_P\rangle_E
\end{eqnarray}
where in the last equality we used the fact that $f(k,q)$ depends only on $k^+$, and that  $q^+=0$.

The matrix element $O_1$ corresponds to the scattering of a single projectile gluon which absorbs and reemits a target gluon, while $O_2$ corresponds to scattering of two distinct projectile gluons,  one absorbing and the other one emitting a gluon from/into the target wave function.

Assuming again rotationally invariant projectile wave function, the average of the product of fields in $O_1$ must be proportional to $\delta_{km}$. We can then simplify \eqref{1ps} to
\begin{eqnarray}\label{o1}
O_1(q)&=&\frac{g^2N_c}{2} \int_{k:k^-<q^-} 
\frac{1}{2k^+}f^i_{lk}(k,q)f^i_{lk}(k,q)
\langle \Psi_P | A^{a\dagger}_m(k^+, \v k)A^a_{m}(k^+, \v k)|\Psi_P\rangle_E\nonumber\\
&=&\frac{g^2N_c}{2} \int_{k:k^-<q^-} 
\frac{1}{(2k^+)^2}f^i_{lk}(k,q)f^i_{lk}(k,q)\, T(k)
\end{eqnarray}
with
\begin{equation}\label{fsq}
[f^{Ti}(k,q)f^i(k,q)]_{st}=
\frac{(2k^+)^2}{  \v q^2}\left[1+\frac{4}{(\frac{k^+}{  \bar q^+}-1)^2}\right]\delta_{st}; \ \ \ \ \ \ \ \ \ 
\end{equation}
Above we have introduced $T$ - the gluon transverse momentum dependent distribution (TMD)
\begin{equation}\hat T(k)\equiv  a^{\dagger a}_i(k) a^a_i(k);   \quad  \qquad T(k)\equiv \langle \Psi_P|\hat T(k)|\Psi_P\rangle_E
\end{equation}
where $ a^\dagger$ and $ a$ are the gluon creation and annihilation operators (rather than the field operators $ A^\dagger$ and $ A$)
and $\hat T$  is the particle number  operator.
Perhaps unsurprisingly, the observable \eqref{1ps} is expressed entirely in terms of the gluon TMD. In fact, in the eikonal limit  
 $\bar q^+\ll k^+$, we have
 \begin{equation}
 [f^{Ti}(k,q)f^i(k,q)]_{st}\rightarrow
\frac{(2k^+)^2}{  \v q^2}
 \end{equation}
 
and  the projectile observable becomes simply the total number of gluons in the projectile wave function - the gluon PDF integrated over the longitudinal momentum fraction. This is natural since in the eikonal limit we assume that all projectile gluons scatter in exactly the same way, an so one simply has to count the total number of gluons in the wave function.

 On the other hand, in the hard scattering limit  \eqref{nbfkl} is sharply 
 peaked at  $k^+\sim\bar q^+$.
 In this limit a good approximation is 
 \begin{equation}\label{fap}
 [f^{Ti}(k,q)f^i(k,q)]_{st}\approx
\frac{(2k^+)^2}{ \v q^2}\frac{1}{\epsilon}\delta(k^+-\bar q^+)\delta_{st}\,.
\end{equation}
for some constant $\epsilon$. This formulae is somewhat impressionistic, since \eqref{fap} is integrable with respect to $k^+$ while the original form \eqref{fsq} is not, which suggests $1/\epsilon\rightarrow\infty$. However this problem is only apparent. We have to remember that in the expression for cross section \eqref{fsq} is multiplied by the target average. This averaging smears the value of $\bar q$, since the target field distribution is not a strict delta function in frequency. For example, assuming that the distribution is sharp in transverse momentum, i.e. $\v q^2$ is fixed, but the distribution in $q^-$ has a (small) width $\Delta q^-$, we find $\epsilon =\frac{\Delta q^-}{\v q^2}$. 
Since $\epsilon$ is determined by the properties of the target distribution, it is convenient to attribute it not to the operator $O_1$ but rather to the target part of the cross section, i.e. $\bar O_1\equiv\epsilon O_1$.
With this factor scaled out, the single particle  part of the scattering amplitude becomes
\begin{equation}\label{dis}
\bar O_1(q)=\frac{g^2N_c}{4\pi}\frac{1}{  q^2} \int_{  \v k:\ \v k^2< \v q^2=2k^+q^-}  
T( k^+=\bar q,\v k)
\end{equation}

We can now make a direct connection with DIS cross section. First, recall that $q^-\approx E$. Now  introduce the standard DIS variables: the total center of mass energy squared $s=2P^+E$;
the momentum transfer $Q^2\equiv \v q^2$, and  the Bjorken $x$ given by  $x\equiv k^+/P^+= Q^2/s$ (at small $x$).
With the natural identification of the gluon PDF as
\begin{equation}\label{pdf}
G(x,Q^2)=\int_{ \v k: \  \v k^2<Q^2}
T( k^+,\v k),
\end{equation}
eq. \eqref{dis} (when substituted into \eq{Sga}) becomes the standard  relation between the scattering cross section and the gluon PDF  in DIS (for a probe that couples to gluons rather than quarks). Note that the transverse resolution scale for the PDF in \eqref{pdf} is given in a natural way by the transverse momentum transfer from the target $Q^2$. 
This value for the transverse resolution scale is expected also from another point of view: $2k^+E$ is  the highest transverse momentum that gluons with longitudinal momentum fraction $x$ can have in the evolved wave function, and thus it aligns nicely with the understanding of $Q^2$ as the transverse resolution scale. In the second paper of this series \cite{second} we discuss in detail this definition of PDF and gluon TMD as well as their frequency evolution. 

The two particle scattering contribution $O_2$ in the hard scattering regime provides a power correction to $O_1$ suppressed by  $Q^2$. In the soft scattering regime, which is dominated by the scattering of gluons with low $x$, the two contributions play equally important roles.

In the subsequent papers in this series we will study in detail the evolution of the cross section and will separately discuss the hard scattering (DGLAP) and soft scattering (BFKL) regimes. 

In the remainder of the present paper our goal is to demonstrate that the BO evolution encompasses the most important aspects of the calculations that lead to the derivation of the NLO JIMWLK equation \cite{LuMu}. Specifically we will examine the one gluon emission amplitude which (among other quantities) was calculated in \cite{LuMu}  and will demonstrate that the perturbative expansion of the evolved wave function derived here reproduces large transverse logarithms that arise in the calculation of this amplitude in \cite{LuMu}.

 \section{Reproducing the JIMWLK large logarithms -- the single gluon amplitude at order $g^3$}
 
 The NLO JIMWLK Hamiltonian in the operator approach was derived in \cite{LuMu}. This calculation was performed in the "+" scheme, i.e. assuming evolution in $k^+$, and was the basis of the derivation of the NLO JIMWLK Hamiltonian. 
 In \cite{LuMu}  the authors calculated the one and two gluon emission amplitudes
 assuming that the gluons has rapidity in the highest rapidity bin of width $\Delta $. Here we will concentrate on the single gluon amplitude, since it contains the large transverse logarithms in the explicit form. We will also adapt the calculation of \cite{LuMu} to the $k^-$ ordering by simply requiring that the width of the the rapidity bin is $\Delta$ in $\ln k^-$ rather than in $\ln k^+$.
 
 The procedure by which \cite{LuMu} deduced the NLO JIMWLK Hamiltonian was the following. The calculation of the amplitude was performed perturbatively to order $g^3$, and the emission probability was then calculated to order $g^4$. The terms in the probability of order $(\Delta )^2$ where then identified as the second iteration of the leading order JIMWLK kernel and discarded. The terms of order $\Delta $ where  fully attributed to the NLO correction to the JIMWLK kernel.  We feel that this interpretation of the perturbative calculation of the probability is somewhat questionable  for the following reason. To perform the actual calculation one had to assume that the rapidity bin is wide enough. This is not a problem since at weak coupling one can have $\Delta \gg 1$, but $\alpha_s\Delta\ll 1$. However  in order to extract the large transverse logarithmic terms one also had to assume $\Delta\gg \ln \frac{\v p^2}{\v k^2}$. In order to derive the differential form of the JIMWLK evolution one must have $\alpha_s\Delta\ll 1$. Under these assumptions the transverse logarithm can never be large, and it is thus unclear how one can consistently take the limit  $\alpha_s\Delta\rightarrow 0$ and still claim to obtain large logarithmic corrections in the kernel.
 
 For this reason here we will not interpret the calculation of \cite{LuMu} as the calculation of JIMWLK kernel, but rather in the simplest way, purely as the calculation of the single gluon emission amplitude to order $g^3$. 
 
 Our goal in this section is to reproduce the results of \cite{LuMu} understood in this simple way, by utilizing explicitly the wave function derived in the present paper.
 
 We denote the  wave-function of gluon modes in the momentum interval $\Delta$ computed in \cite{LuMu} by $|\Psi_{NLO} \rangle$. The object of interest in this section is the one gluon amplitude to order $g^3$, $\langle \v k,b,i|\Psi_{NLO} \rangle$, where the
momentum of the gluon is $k=(k^+,\v k)$, its color is $b$ and polarization $i$. In the next subsection we will briefly review the results of \cite{LuMu} 
and demonstrate how large transverse logarithms arise in this calculation. 

As mentioned above the calculation of \cite{LuMu} is performed in "+" scheme, which in the present context means simply that the kinematic range defined by $\Delta$ is
\begin{equation}
\label{+window}
 \Lambda_+e^{-\Delta}<k^+<\Lambda_+
\end{equation}
 where $\Lambda_+$ is the separation scale scale, in the sense that the effect of all modes with $k^+>\Lambda_+$ is contained in the color charge density $\rho$. As we will see below it is easy to infer from this calculation the result in the BO scheme, where the rapidity interval $\Delta$ defines the frequencies of the emitted gluons, i.e. 
 \begin{equation}\label{-window}
 Ee^{\Delta}>k^->E
 \end{equation}
  Note that in this section we do not assume that $\Delta$ is small.
 

\subsection{\cite{LuMu} revisited: $\langle  {\bf k}|\Psi_{NLO} \rangle$ from "+" to "-" scheme}

The result for the amplitude in  \cite{LuMu}  is presented as \footnote{Where the one particle state is normalized as $|k,i\rangle=\frac{a^{i\dagger}(k^+, \v k)}{(2\pi)^{3/2}}|0 \rangle$}
\begin{equation}\label{F2}
{1\over \sqrt{2k^+}}\,\langle \v k,b,i|\Psi_{NLO} \rangle= F^1_i\rho^b(-k) +  \int_p
\left[i F^2_i f^{abc} \{\rho^c(p-k),\rho^a(-p)\}_+ + F^3_i \rho^b(-k) \rho^a(p)\rho^a(-p)\right]
\end{equation}
To calculate the probability of gluon emission to order $g^4$ we need to multiply the amplitude $\langle \v k,b,i|\Psi_{NLO} \rangle$   by the single gluon amplitude at order $g$. 
The factor $1/\sqrt{2k^+}$ in \eqref{F2} to all intents and purposes achieves this, since the $k^+$ dependence  of the leading order amplitude is given precisely by this factor. The additional transverse momentum dependent factors do not affect the large transverse logarithms since only the longitudinal momenta are integrated over in the calculation of \cite{LuMu}.

Here, as is common in the $+$ scheme, $\rho^a(p)$ is taken to be $p^+$ independent.
Furthermore,  $\rho^a$ are quantum operators, and therefore do not commute with each other. Their relative ordering is therefore important.

The terms linear in $\rho$ contain virtual corrections. They do contain a large transverse logarithm associated with DGLAP splittings, as discussed at length in \cite{ourdglap}. In fact this term is related to the BO evolution of the gluon TMD, which we will discuss in detail in \cite{second}. Here therefore we will not dwell on this term.


The  $\rho^3$  term is simple and we will discuss it later. 

For now we concentrate on  the $\rho^2$ terms. 
The result of \cite{LuMu} for these terms is
\begin{equation}\label{fb}
F^2_i(k,p)=\frac{1}{2(2\pi)^{3/2}}\frac{g^3}{(2\pi)^3} \sum_{n=3}^7\psi^n_i(k,p)
\end{equation}
with
\begin{eqnarray}\label{psii}
\psi^3_i&=&\frac{k^++p^+}{(k^+-p^+)^2}\frac{1}{p^+}\frac{\v p_i}{\v k^2\v p^2}\nonumber\\
\psi^4_i&=&-\frac{k^+-p^+}{(k^++p^+)^2}\frac{k^+}{p^+}\frac{\v p_i}{\v k^2[k^+\v p^2+p^+\v k^2]}\nonumber\\
\psi^5_i&=&\theta(p^+-k^+)\frac{1}{\v k^2\v p^2}\frac{k^+}{p^+}\frac{1}{[k^+(\v k-\v p)^2+(p^+-k^+)\v k^2]}\times\nonumber\\
&&\Bigg\{\Big[-\frac{k^++p^+}{(k^+-p^+)^2}(\v k-\v p)^2+\frac{1}{k^+-p^+}(\v k^2-\v p^2)-2\frac{1}{p^+}\v p^2\Big]\v p_i+2\Big[\frac{\v p^2}{p^+}-\frac{\v p(\v k-\v p)}{k^+}\Big]\v k_i\Bigg\}\nonumber\\
\psi^6_i&=&\theta(k^+-p^+)\frac{1}{\v  k^2\v p^2}\frac{1}{[p^+(\v k-\v p)^2+(k^+-p^+)\v p^2]}\times\nonumber\\
&&\Bigg\{\Big[-\frac{k^++p^+}{(k^+-p^+)^2}(\v k-\v p)^2+\frac{1}{k^+-p^+}(\v k^2-\v p^2)-2\frac{1}{p^+}\v p^2\Big]\v p_i+2\Big[\frac{\v p^2}{p^+}-\frac{\v p(\v k-\v p)}{k^+}\Big]\v k_i\Bigg\}\nonumber
\\
\psi^7_i&=&2\frac{1}{\v k^2}\frac{1}{p^+}\frac{1}{k^+\v p^2+p^+\v k^2}\v k_i 
\end{eqnarray}
where we did a little bit of algebra to simplify the original expressions.

Here we recorded $\psi^7$ for completeness, but this term is irrelevant. It arises due to reordering of the color charge density factors in the $\rho^3$ term. In \cite{LuMu} a particular ordering of the factors was chosen, and this lead to appearance of $\psi^7$ in the form of \eqref{psii}. If, on the other hand one keeps the $\rho^3$ term in the exact form it directly arises in the diagrammatic calculation, no $\psi^7$ is present.  This will be our strategy here.

The large transverse logarithms originate from integrations over the momentum $p$ in \eqref{F2}. In this respect we note that \eqref{psii} does not assume the choice of $k^+$ or $k^-$ scheme. This choice comes into fore only when integrating over momenta by imposing the limits of integration.


The amplitude (\ref{F2}) involves integration over $p^+$, and in addition 
the NLO JIMWLK derivation of \cite{LuMu} involves integration  over $k^+$.  Both $p^+$ and $k^+$ are integrated over the interval \eqref{+window}.


We define amplitudes integrated over the longitudinal momenta 
\begin{equation}
\bar\psi_i^n\equiv \int_{k^+,p^+}\psi_i^n\,;\qquad n=3...7
\end{equation}  
In the "+" scheme employed in \cite{LuMu} the integrations over  the longitudinal momenta are taken over the interval  \eqref{+window}. %
With some algebra (see Appendix \ref{psisec}), we can write,
\begin{eqnarray}\label{psiABC}
\bar\psi^5_i+\bar\psi^3_i+\bar\psi^4_i+\bar\psi^6_i&=&-(2B^+-A^+)\frac{(\v k-\v p)_i}{(\v k-\v p)^2}
+[B^+-A^+-C^+]\frac{\v p\cdot \v k\v k_i-\v k^2\v p_i}{\v k^2(\v k-\v p)^2}\nonumber \\
&+&[B^++A^+-C^+]\frac{\v p\cdot(\v k-\v p)}{(\v k-\v p)^2}\frac{\v k_i}{\v k^2}\,;\qquad\qquad
\bar\psi^7_i=2B^+\frac{\v k_i}{\v k^2}
\end{eqnarray}
where $A^+,B^+,C^+$ are given by,
\begin{equation}
\begin{split}
A^+=&\int_{p^+,k^+}\frac{1}{k^++p^+}\frac{1}{k^+\v p^2+p^+\v k^2} \\
 =& \frac{1}{\v p^2}\left({\Delta}-\frac{1}{2}\ln \frac{\v p^2}{\v k^2}\right)\ln\frac{\v p^2}{\v k^2}\theta[{\Delta}-\ln\frac{\v p^2}{\v k^2}]+\frac{1}{2}\frac{1}{\v p^2}{\Delta}^2\theta[\ln\frac{\v p^2}{\v k^2}-{\Delta}]\\
B^+=&\int_{p^+,k^+}\frac{1}{p^+}\frac{1}{k^+\v p^2+p^+\v k^2}\\
 =& \frac{1}{\v p^2}\Big\{\Big[\frac{1}{2}{\Delta}^2+{\Delta}\ln \frac{\v p^2}{\v k^2}-\frac{1}{2}\ln^2 \frac{\v p^2}{\v k^2}\Big]\theta[{\Delta}-\ln\frac{\v p^2}{\v k^2} ]+{\Delta}^2\theta[\ln\frac{\v p^2}{\v k^2}-{\Delta}]\Big\}\\
C^+=&\int_{p^+,k^+}\theta(k^+-p^+)\frac{1}{p^+}\frac{1}{[p^+(\v k-\v p)^2+(k^+-p^+)\v p^2]}\\
 =& \frac{1}{2\v p^2}{\Delta}^2
 \end{split}
\end{equation}

This leads to,
\begin{eqnarray}\label{ABC}
B^+-A^+-C^+&=& 0\\
B^++A^+-C^+&=& 2\left[{\Delta}\ln\frac{\v p^2}{\v k^2}-\frac{1}{2}\ln^2\frac{\v p^2}{\v k^2}\right]\theta[{\Delta}-\ln\frac{\v p^2}{\v k^2} ]+{\Delta}^2\theta[\ln\frac{\v p^2}{\v k^2}-{\Delta}]\nonumber\\
2B^+-A^+&=&\frac{1}{\v p^2}\left [{\Delta}^2+{\Delta}\ln\frac{\v p^2}{\v k^2}-\frac{1}{2}\ln^2\frac{\v p^2}{\v k^2}\right]\theta[{\Delta}-\ln\frac{\v p^2}{\v k^2} ]+ \frac{3}{2}\frac{1}{\v p^2}{\Delta}^2\theta[\ln\frac{p^2}{k^2}-{\Delta}]\nonumber
\end{eqnarray}
For  $\Delta\gg \ln\frac{\v p^2}{\v k^2}$ (as  assumed in \cite{LuMu}) the last terms in both expressions in \eqref{ABC} vanish. In the same regime the terms involving $\ln^2\frac{\v p^2}{\v k^2}$ are negligible and the expression \eqref{psiABC}  becomes
\begin{eqnarray}\label{psi16}
\Big[\bar\psi^5_i+\bar\psi^3_i+\bar\psi^4_i+\bar\psi^6_i\Big]
&=&\left[\frac{1}{\v p^2}{\Delta}^2+\frac{1}{\v p^2}{\Delta}\ln\frac{\v p^2}{\v k^2}\right]\frac{(\v p-\v k)_i}{(\v p-\v k)^2}
+{2\over \v p^2}{\Delta}\ln\frac{\v p^2}{\v 	k^2}\ \frac{\v p\cdot(\v k-\v p)}{(\v k-\v p)^2}\frac{\v k_i}{\v k^2}
\end{eqnarray}
The large logarithm in these expressions is $\ln\frac{\v p^2}{\v k^2}$ as the integral over $p$ runs over the whole range including  $\v p^2\gg \v k^2$.
We recall that in \eqref{F2}, the combination (\ref{psi16}) is convoluted with the $\rho(\v p)\rho(\v k-\v p)$, which is 
antisymmetric  (because of the color factor) under the exchange of   $\v p\rightarrow \v k-\v p$. The last term in (\ref{psi16}) is  symmetric under this change of variables, except the $\ln \v p^2/\v k^2$. Effectively, this means that instead of  
 $\ln \v p^2/\v k^2$ we can write ${1\over 2} \ln \frac{\v p^2}{(\v p-\v k)^2}$ so that there is no large logarithm here in the limit
 $\v p^2\gg \v k^2$. Thus
 \begin{eqnarray}\label{psi17}
\Big[\bar\psi^5_i+\bar\psi^3_i+\bar\psi^4_i+\bar\psi^6_i\Big]
&\simeq&\frac{1}{\v p^2}\left[{\Delta}^2+{\Delta}\ln\frac{\v p^2}{\v k^2}\right]\frac{(\v p-\v k)_i}{(\v p-\v k)^2}
\end{eqnarray}


We can now use these results to find the expression for the probability in the $-$ scheme, which is what we will compare with the BO calculation. Define

\begin{eqnarray}
A^-&\equiv&\int_{p^+,k^+}\frac{1}{k^-+p^-}\frac{1}{k^-\v p^2+p^-\v k^2}\nonumber\\
B^-&\equiv&\int_{p^+,k^+}\frac{1}{\v p^2}\frac{1}{k^-}\frac{1}{k^-+p^-}\nonumber\\
C^-&\equiv&\int_{p^+,k^+}\theta\left(p^--k^-\frac{\v p^2}{\v k^2}\right)\frac{1}{\v p^2}\frac{1}{p^-}\frac{1}{k^-}
\end{eqnarray}
where now the integration range is given by \eqref{-window}.

The integrals can be readily performed with the result 
\begin{eqnarray}\label{-ABC}
A^-&=&\frac{1}{\v p^2-\v k^2}\left[\Delta\ln \frac{\v p^2}{\v k^2}-\frac{1}{2}\ln^2 \frac{\v p^2}{\v k^2}\right]\theta\left(\Delta-\ln\frac{\v p^2}{\v k^2}\right)+\frac{1}{2\v p^2}\Delta^2\theta\left(\ln\frac{\v p^2}{\v k^2}-\Delta\right)
\nonumber \\
B^-&=&\frac{1}{2\v p^2}\Delta^2\nonumber \\
C^-&=&\frac{1}{2\v p^2}\Big[\Delta-\ln\frac{\v p^2}{\v k^2}\Big]^2\theta\left(\Delta-\ln\frac{\v p^2}{\v k^2}\right)
\end{eqnarray}
For $\v p^2\gg \v k^2$,
\begin{eqnarray}
&&B^--A^--C^-=0\\
&&2B^--A^-=\frac{1}{2\v p^2}\Delta^2+\frac{1}{2\v p^2}\Big[\Delta-\ln\frac{\v p^2}{\v k^2}\Big]^2\theta\left(\Delta-\ln\frac{\v p^2}{\v k^2}\right)\nonumber\\
&&B^-+A^--C^-=\frac{2}{\v p^2}\left[\Delta\ln \frac{\v p^2}{\v k^2}-\frac{1}{2}\ln^2 \frac{\v p^2}{\v k^2}\right]\theta\left(\Delta-\ln\frac{\v p^2}{\v k^2}\right)+\frac{1}{\v p^2}\Delta^2\theta\left(\ln\frac{\v p^2}{\v k^2}-\Delta\right)\nonumber
\end{eqnarray}
The relation \eqref{psiABC} is purely algebraic, and therefore it holds also in $-$ scheme, so that now we have 

\begin{eqnarray}\label{sumpsi}
\Big[\bar\psi^5_i+\bar\psi^3_i+\bar\psi^4_i+\bar\psi^6_i\Big]&=&\left[\frac{1}{2\v p^2}\Delta^2+\frac{1}{2\v p^2}\Big[\Delta-\ln\frac{\v p^2}{\v k^2}\Big]^2\theta\left(\Delta-\ln\frac{\v p^2}{\v k^2}\right)\right]\frac{(\v p-\v k)_i}{(\v p-\v k)^2}\\
&+&\left\{\frac{2}{\v p^2}\left[\Delta\ln \frac{\v p^2}{\v k^2}-\frac{1}{2}\ln^2 \frac{\v p^2}{\v k^2}\right]\theta\left(\Delta-\ln\frac{\v p^2}{\v k^2}\right)+\frac{1}{\v p^2}\Delta^2\theta\left(\ln\frac{\v p^2}{\v k^2}-\Delta\right)\right\}\frac{\v p\cdot(\v k-\v p)}{(\v k-\v p)^2}\frac{\v k_i}{\v k^2}\nonumber\\
&\approx&\frac{1}{\v p^2}\Delta^2\frac{(\v p-\v k)_i}{(\v p-\v k)^2}-\frac{1}{\v p^2}\Delta\ln\frac{\v p^2}{\v k^2}\left[\frac{(\v p-\v k)_i}{(\v p-\v k)^2}-\frac{2\v p\cdot(\v k-\v p)}{(\v k-\v p)^2}\frac{\v k_i}{\v k^2}\right]\nonumber\\
&\simeq& \frac{1}{\v p^2}\left[\Delta^2-\Delta\ln\frac{\v p^2}{\v k^2}\right] \,\frac{(\v p-\v k)_i}{(\v p-\v k)^2}
\nonumber
\end{eqnarray}
where the last approximate equality is valid for $\Delta\gg\ln\frac{\v p^2}{\v k^2}\gg 1$ (here we again used 
 the argument, which lead us to \eqref{psi17}). 

\subsubsection*{Comments}
To conclude this review of \cite{LuMu} we offer a couple of comments which are not directly related to our goal in this section.

The first point is the putative derivation of the NLO JIMWLK Hamiltonian from the above calculations.
In \cite{LuMu},  the $\Delta^2$ term in \eqref{psi16} was associated with the second iteration of the leading order Hamiltonain, while $\Delta \ln\frac{\v p^2}{\v k^2}$ was regarded as genuine  NLO result and attributed directly to the NLO correction of the Hamiltonian by taking formally the limit $\Delta\rightarrow 0$. The terms involving $\ln\frac{\v p^2}{\v k^2}$ then may give rise to  large corrections upon integration over $p$. However the result \eqref{psi16} is only valid for $\Delta>\ln\frac{\v p^2}{\v k^2}$. For the opposite kinematics, $\Delta<\ln\frac{\v p^2}{\v k^2}$
, the last terms in \eq{ABC} take over and there seem to be no large transverse logarithms at all. The situation is therefore somewhat perplexing. On one hand in order to see large logarithms we must perform the integration over large rapidity interval $\Delta$. On the other hand in order to derive the correction to the evolution Hamiltonian we need to take the formal limit $\Delta\rightarrow 0$, in which limit the transverse logarithms disappear. This perhaps explains our attitude in the present paper to consider the calculation of \cite{LuMu} as just the calculation of the amplitude at order $g^3$.


Our second point is the question
how do we see that the transverse logarithm has a large effect in the amplitude?
  Clearly, in order for $\ln \frac{\v p^2}{\v k^2}$ to give a large contribution,  the region of the phase space $\v p\gg \v k$ must give a significant contribution in the integral over $\v p$ in \eq{F2}. Yet, the kernel of $\v p$ integration drives the integral towards small $\v p$, that is the integral is seemingly dominated by $\v p\simeq \v k$, where the logarithm is small. This is explicit for the first term in \eqref{psi16} and \eqref{sumpsi}. In the second term we note that the factor $\frac{\v p\cdot(\v k-\v p)}{\v p^2(\v k-\v p)^2}$ is symmetric under $\v p\rightarrow \v k-\v p$, and therefore  for large $\v p$ it vanishes faster than $1/\v p^2$. 
  
  One has to remember however,  that  in \eq{F2}, the probability  $F^2_i$ is multiplied by the product of color charge densities. With a simple perturbative assumption  that $\langle\rho(p)\rho(-p)\rangle$ is independent of $p$, it is indeed  true that the transverse logarithms in \eqref{psi16} and \eqref{sumpsi} do not lead to large corrections.  
 But this assumption is too naive. We know that for a globally color singlet projectile state, $\rho(p)$ has to vanish for small $p$. For example, if the projectile is described by the (generalized) MV model, color neutrality requires  $\langle \rho(p)\rho(-p)\rangle \sim \mu^2\v p^2/(\v p^2+Q_s^2)$.  Here the saturation momentum $Q_s$ appears as the color neutralization scale in the projectile. 
Also if the projectile is a dipole of size $R$, the color charge density vanishes for $|\v p|<1/R$. The latter example is pertinent for numerical work based on the BK equation. 

Let us denote the non-perturbative scale, below which $\rho$ vanishes by $Q_s$.  Now two distinct regimes arise. For $|\v k|\gg Q_s$ indeed the transverse logarithm is unimportant as $|\v p|$ is driven to small values by the integrand. On the other hand for $|\v k|<Q_s$ the situation is different. Here small values of $p$ are cutoff by the color charge density, and the integration over $p$ is {\it de facto}  limited from below by $p\sim Q_s$. The integral over $p$ therefore will indeed produce a large logarithm $\ln Q_s/|\v k|$.  This NLO result  must be compared with LO, which is proportional to $\langle \rho(k)\rho(-k)\rangle$, and is very small  for $Q_s\gg |\v k|$. Thus if the saturation momentum of the projectile is large enough there is a sizable range of transverse momenta $k$ for which the NLO correction to the amplitude is enhanced  by the large logarithm. In the context of BK equation for example, this large correction appears for momenta of emitted gluon smaller than the inverse size of the parent dipole, which takes the role of $1/Q_s$. In coordinate space this corresponds to  configurations where the daughter gluon is emitted at distances much larger than the size of the parent dipole. 


 \subsection{$\langle {\bf k}|\Psi_{P} \rangle$ -- reproducing  the large logarithms in BO scheme.}

 Our goal here is to start from the wave function $|\Psi_P\rangle_E$ (\ref{psiel}) and 
 calculate the amplitude $
 \langle \v k|\Psi_P\rangle_E
 $
 for a single gluon state at order $g^3$. The  frequency of the gluon is assumed to be in the window $E_0<k^-<E$ with $\Delta=\ln\frac{E}{E_0}$.
 We will organize the calculation as expansion in powers of $\rho$ in order to facilitate comparison with \cite{LuMu}. The single gluon amplitude contains terms of order $\rho^3$, $\rho^2$ and $\rho$. 
We are going to split the charge density $\rho^a(p^+,p)$
in  (\ref{rho+}) as
 \begin{eqnarray}\label{rho++}
\rho^a(p^+,\v p)&=& \rho^a(p)\,+\,\delta \rho^a( p); \nonumber \\
 \rho^a(p)&=&
-2i
f^{abc}\int_{q^-<E_0 } \,q^+\,A^{b\dagger}_{j}(q^+,\v q+\v p) A^c_{j}(q^+,\v q)\,;\nonumber \\
 \delta\rho^a( p)&=&
-2i
f^{abc}\int_{E_0<q^-<p^- } \,q^+\,A^{b\dagger}_{j}(q^+,\v q+\v p) A^c_{j}(q^+,\v q)
\end{eqnarray}
The the first term is made entirely of the gluons whose frequencies are  below "the window".  
We will call them far away valence modes (FAVM). This term 
 corresponds to $\rho$ of \cite{LuMu}
The gluons making up $\delta\rho$ themselves live inside the window. If the width $\Delta$ is not small at NLO these gluons can themselves emit other gluons in the window, and thus their contribution to the charge density has to be taken into account.

\subsubsection{The order $\rho^3$ terms.}
It is obvious that this part of the amplitude does not involve $G^{DGLAP}$, and comes entirely
from the third power of $G^{BFKL}$ term where all factors $\rho$ are made of FAVMs.
\begin{eqnarray}
&&|\Psi_{3}\rangle\equiv |\Psi_P\rangle_E|_{\rho\rho\rho}=8g^3\int_{p^-<q^-<m^-}\Big[\rho ^c(-m)\frac{\v m_k}{\v m^2}A_k^{\dagger c}(m^+,\v m)\rho ^{*b}(-q)\frac{\v q_j}{\v q^2}A_j^b(q^+,\v q)\rho ^a(-p)\frac{\v p_i}{\v p^2}A_i^{\dagger a}(p^+,\v p)\nonumber\\
&&+\rho ^{*c}(-m)\frac{\v m_k}{\v m^2}A_k^ c(m^+,\v m)\rho ^b(-q)\frac{\v q_j}{\v q^2}A_j^{\dagger b}(q^+,\v q)\rho ^a(-p)\frac{\v p_i}{\v p^2}A_i^{\dagger a}(p^+,\v p)\Big]  |0\rangle_F \nonumber \\
&&=8g^3\int_{p^-<q^-}\Big[\frac{1}{2p^+}\frac{\v q_i}{\v q^2\v p^2}\rho ^a(-q)\rho ^{*b}(-p)\rho ^b(-p)A_i^{\dagger a}(q^+,\v q)+\frac{1}{2q^+}\frac{\v p_i}{\v q^2\v p^2}\rho ^a(-p)\rho ^{*b}(-q)\rho ^b(-q)A_i^{\dagger a}(p^+,p)\Big] | 0\rangle_F\nonumber\\
&&=8g^3\int_{p,q}\frac{1}{2p^+}\frac{\v q_i}{\v q^2\v p^2}\rho ^a(-q)\rho ^{*b}(-p)\rho ^b(-p)A_i^{\dagger a}(q^+,\v q) | 0\rangle_F
\end{eqnarray}
The frequencies of all the slow field modes here are within the window.

To get the last equality we have renamed some of the momenta. 
In the expressions above integrations over all momenta are implied. This yields (multiplying by $1/\sqrt{2k^+}$)  
\beq
\frac{1}{\sqrt{2k^+}}\langle \v k |\Psi_{3}\rangle
=\frac{4g^3}{(2\pi)^{3/2} [2k^+]}\int_{E_0}^{E} {dp^-\over 2\pi}\int {d^2\v p\over (2\pi)^2} \frac{1}{p^+}\frac{\v k_i}{\v k^2\v p^2}\rho ^a(-k)\rho ^{*b}(-p)\rho ^b(-p)
\eeq
In the last expression the integral over $p^-$ extends over the whole allowed range, $E_0<p^-<E$ and is not ordered relative to $k^-$.
To compare with \cite{LuMu}  we need to order the factors of $\rho$ in the same way as it was done in \cite{LuMu}. To achieve this we have to commute the factor $\rho^a(-k)$ through $\rho^{*b}(-p)\rho^b(-p)$. This commutator is nontrivial and produces the term proportional to $\rho^2$, which reproduces exactly the amplitude $\psi^7$ from \cite{LuMu}. The remainder of the terms, which are all of order $\rho^3$ also match the calculation of \cite{LuMu}.

\subsubsection{The order $\rho^2$ terms.}

We first note that  $G^{DGLAP}$ creates or annihilates two gluons in the window. Therefore it does not contribute to $\rho^2$ terms in the single gluon amplitude, since each factor of $\rho$ is associated with one eikonal vertex which creates a single gluon. Thus this amplitude comes entirely from expansion of the wave function in powers of $G^{BFKL}$ (\ref{gbfkl}). 

As briefly mentioned earlier, in order to recover all the large logarithms of  \cite{LuMu} it is important to take into account  the soft mode $\alpha$ 
introduced in Section 2. We dropped this mode in our earlier discussion since we were interested in calculating the leading order wave function of the modes with highest available frequency. The soft mode by definition has the smallest frequency than any other mode it interacts with, and thus by definition it does not live in the last frequency bin of the evolved wave function. Here however we are considering production of one gluon in a large rapidity interval. Thus it is entirely possible to produce such a gluon whose frequency is somewhere inside the interval, far away from  its edge.

We therefore restore it here by reintroducing the covariant derivative $P_i$ in the $\alpha$ background (\ref{Palpha}) into the eikonal production vertex:
\begin{eqnarray}\label{gbfkl1}
G^{BFKL}(p^+,\v p^2)&=&2g \left[A_{i}^{\dagger a}(p) P_j [Q^{-1} ]_{ij}^{ab}
\rho^b(p^+,-\v p)+A_{i}^{ a}(p)P_j [Q^{-1} ]_{ij}^{ab}  \rho^{\dagger b}(p^+,-\v p)\right]
\end{eqnarray}
The matrix $Q$ is defined as $Q_{ij}^{ab}\equiv (\v P^{2}\delta_{ij}+[\v P_i,\v P_j] )^{ab}$\footnote{In principle one might also worry about the contribution from $\Delta H$ (\ref{dH}). This contribution, however, cancels in 
 a very subtle way discussed in the Appendix \ref{soft}.}. The reason for appearance of $Q^{-1}$ is that the soft background $\alpha$ modifies the kinetic term of the other, faster modes in the light cone Hamiltonian and thus modifies the energy denominator which enters in $G^{BFKL}$.
This expression  in general is highly non-linear in $\alpha$, however since we only need the amplitude to order $g^3$ it is sufficient to keep only terms linear in $\alpha$. 
Hence, we can compute the inverse of $Q$ perturbatively
 \begin{equation}
 [Q^{-1} ]_{ij}^{ab}= \frac{1}{\v p^{2}}(\delta^{ab} \delta_{ij}-2  i gf^{abc}  {\v p_k\alpha_k^c\over \v p^2} \delta_{ij}
 + i gf^{abc}  {\v p_i\alpha_j^c\over \v p^2} + i gf^{abc}  {\v p_j\alpha_i^c\over \v p^2})
 \end{equation}
  \begin{equation}
 \v P_j[Q^{-1} ]_{ij}^{ab}= \frac{1}{\v p^{2}}\left(\delta^{ab} \v p_{i}-  i gf^{abc}  {\v p_i\v p_{k}\alpha_k^c\over \v p^2} 
 + 2 i gf^{abc}  \alpha_i^c\right)=\frac{1}{\v p^2}\left(\delta^{ab} \v p_{i} +i gf^{abc} \left[ {\v p_i\v p_{k}\over \v p^2}
 -2  ({\v p_i\v p_{k}\over \v p^2} -
  \delta_{ik})\right]\alpha_k^c\right)
 \end{equation}
 Omitting the magnetic term (transverse projector), which does not contribute to the pertinent order in perturbation theory, 
 the final expression for the wave function valid to order $g^3$ reads
\begin{eqnarray}\label{wfun1}
|\Psi_P\rangle
&=&{\cal P}\exp\left\{ 2\int_{ p:\ E_0<p^-<E}\frac{\v p_i}{\v p^2}[A_i^{\dagger a}(p) \rho^b(p^+,-\v p)
+A_i^a(p)\rho^{\dagger b}(p^+,-\v p) ]\right. \\
&\times&\left.\left[g\delta^{ab}+ ig^2f^{abc}\frac{\v p_j}{\v p^2}
\int_{k:\ k^+\ll p^+; k^-\ll p^-}\Big[\alpha_j^{\dagger c}(-k)+\alpha^c_j(k)\Big]\right]\right\}|0\rangle_F\nonumber
\end{eqnarray}


To order $g^3$,  there are two contributions to the one gluon amplitude. The first one 
comes from the third order expansion of the $O(g)$ term  in \eqref{wfun1} (the term proportional to $\delta^{ab}$),
keeping  second order contribution in $\rho$  (the FAVMs)  and a single power of  $\delta\rho$. 
This contribution will be referred to as  eikonal.
The second contribution is due to  the product of a single eikonal vertex and one vertex involving the soft field $\alpha$ (the second term in \eqref{wfun1}) where the charge density in both terms is taken to be $\rho$. We will refer to this term as "the soft contribution".

\subsubsection*{The eikonal contribution}
The $O(g^3)$ contribution to the wave function obtained by keeping only eikonal vertices is 
\begin{eqnarray}
|\Psi_{eik}\rangle&=&
8g^3\int_{E_0<p^-<q^-<m^-<E}\Big\{\frac{\v m_k}{\v m^2}\frac{\v q_j}{\v q^2}\frac{\v p_i}{\v p^2}\left[\delta\rho ^c(-m)\rho ^{*b}(-q)\rho ^a(-p)+\rho^c(-m)\delta\rho ^{*b}(-q)\rho ^a(-p)\right] \nonumber\\
&\times& A_k^{\dagger c}(m^+,\v m)A_j^b(q^+,\v q)A_i^{\dagger a}(p^+,\v p) \\
&&\hspace{-1cm}+\frac{\v m_k}{\v m^2}\frac{\v q_j}{\v q^2}\frac{\v p_i}{\v p^2}\left[\delta\rho ^{*c}(-m)\rho ^b(-q)\rho ^a(-p)+\rho ^{*c}(-m)\delta\rho ^b(-q)\rho ^a(-p)\right]A_k^ c(m^+,\v m)A_j^{\dagger b}(q^+,\v q)A_i^{\dagger a}(p^+,\v p)\Big\} | 0\rangle_F\nonumber
\end{eqnarray}
The first two terms vanish, since here $A$ annihilates the vacuum unless $q^+=p^+$, but if $q^+=p^+$ the charge density $\delta\rho(q)$ acts on a state that does not contain gluons that constitute $\delta\rho(q)$ and thus annihilates it.
In the second line the operator $A(m)$ has to annihilate the gluon created by $A^\dagger(q)$, since otherwise $m^+=p^+$ and the integration range for $q^+$ shrinks to zero. So we have
\begin{equation}
|\Psi_{eik}\rangle=8g^3\int_{p^-<q^-}\frac{1}{2q^+}\frac{1}{\v q^2}\frac{\v p_i}{\v p^2}\left[\delta\rho ^{*b}(-q)\rho ^b(-q)\rho ^a(-p)+\rho ^{*b}(-q)\delta\rho ^b(-q)\rho ^a(-p)\right]A_i^{\dagger a}(p^+,\v p) | 0\rangle_F
\end{equation}
Here and in the rest of this section it is implied that all momentum integrals are over the range \eqref{-window} although we do not indicate this explicitly.
Contracting $\delta\rho$ with $A^\dagger(p)$,
we get
\begin{equation}
|\Psi_{eik}\rangle=-i4g^3\int_{p^-<q^-; p^+<  q^+}\frac{1}{q^+}\frac{1}{\v q^2}\frac{(\v p-\v q)_i}{(\v p-\v q)^2}f^{abc}\rho ^a(-p+q)\rho ^b(-q)A^{\dagger c}_i(p^+,\v p)\v| 0\rangle_F
\end{equation}
Thus (multiplying the amplitude by $1/{\sqrt{2k^+}}$ for direct comparison with \cite{LuMu})
\begin{eqnarray}\label{psi1}
\psi^a_{eik}(k)\equiv\frac{1}{\sqrt{2k^+}}\langle \v k|\Psi_{eik}\rangle
&=&i\frac{4g^3}{(2\pi)^{3/2}[2k^+]}\int_{k^-<p^-;   k^+>p^+}\frac{1}{p^+}\frac{1}{\v p^2}\frac{(\v p-\v k)_i}{(\v p-\v k)^2}f^{abc}\rho ^b(-k+p)\rho ^c(-p)\\
&=&i\frac{4g^3}{(2\pi)^{3/2}[2k^+]}\left[\int_{k^-<p^-}-\int_{k^-<p^-; p^+>  k^+}\right]\frac{1}{p^+}\frac{1}{\v p^2}\frac{(\v p-\v k)_i}{(\v p-\v k)^2}f^{abc}\rho ^b(-k+p)\rho ^c(-p)\nonumber \end{eqnarray}
which is simply the eikonal result with momenta of the real ($k$) and virtual ($p$) gluons ordered with respect to both rapidity, and longitudinal momentum.

The range of integration in this term is simply understood in terms of rapidity ordering. 
Here gluon $p$ with the highest frequency is eikonally emitted from gluon $k-p$ and then reabsorbed by a valence charge, while the gluon $k$ is produced in the wave function. Since the emission vertex is eikonal,  we are assured that $p^+\ll k^+\approx (k^+-p^+)$. 
On the other hand due to rapidity ($k^-$) ordering we also have $p^-\gg k^-$ and $p^-\gg (k-p)^-$. However we do not have an {\it a priori}  natural ordering between $k^-$ and $(k-p)^-$. Thus when the rapidity interval is wide enough, 
there are 
two distinct contributions.

A) The first contribution comes from the region where $k^->(k-p)^-$. In this range we have $\v k^2>(\v k-\v p)^2$, and thus by momentum conservation $\v p^2\sim \v k^2$. In this range 
 of integration   $p^-$ is limited from below by $k^-$, since this is the highest of the two rapidities of the gluons constituting $\delta\rho$ in the eikonal vertex. For this contribution the longitudinal momentum restriction $p^+\ll k^+$ is subsumed by $k^-\ll p^-$ and is therefore redundant. The integration over $p^-$ is therefore bounded from below by $k^-$.

B) The second contribution comes from the region $(k-p)^->k^-$. In this region  $\v k^2<(\v k-\v p)^2$. This contribution therefore contains the kinematic region $\v p^2\gg \v k^2$. In this region the restriction $p^+\ll k^+$ genuinely restricts the phase space, and it translates on the bound on frequencies   $p^->\frac{\v p^2}{\v k^2}k^-$ rather than $p^->k^-$. Note that this phase space constraint is simply equivalent to $p^->(k-p)^-$.  Thus this is again nothing but imposing the rapidity ordering that requires that the frequency of the gluon $p$ is higher than that of both gluons $k$ and 
$k-p$. 
 In our wave function this is a direct consequence of the $p^-$ ordering. 

We learn from this little digression that when calculating the emission probability of the highest frequency gluon in the "last" frequency bin we cannot restrict the emitting gluons in both, amplitude and conjugate amplitude to belong to the "next-to-last"  bin. One of these gluons can have a frequency significantly smaller than the other, and this lower frequency should be integrated over all allowed values below $E_0$.

\subsubsection*{The soft contribution}

Now let us consider the contribution of the vertex containing a soft gluon $\alpha$. 
Since this vertex in \eqref{wfun1} is of order $g^2$, only one insertion of this vertex accompanied by one eikonal vertex 
that involves FAVM needs to be considered to the order we are working.
There are only two terms in the expansion of the wave function to this order that can potentially create a one gluon state.
Schematically we denote them as $(g\rho A)(g^2\rho \alpha^\dagger A^\dagger)$and $(g^2\rho \alpha^\dagger A)(g\rho A^\dagger)$. The rest of the combinations of the creation and annihilation operators cannot lead to  non-vanishing contractions due to mismatch in frequencies. 
\begin{eqnarray}
&&|\Psi_{soft}\rangle 
=i8g^3\int_{p^->q^->k^-;\  q^+>k^+}\frac{\v p_i\v q_j(\v q+\v k)_l}{\v p^2\v q^2(\v q+\v k)^2}f^{abc}\rho ^{*d}(-p)\rho ^b(-q-k)A^d_i(p)\alpha^{\dagger c}_l(k)A^{\dagger a}_j(q)|0\rangle_F\nonumber\\
&&-i8g^3\int_{p^->q^-;\ p^->k^-;\  p^+>k^+} \frac{(\v p+\v k)_j\v p_i\v q_l}{(\v p+\v k)^2\v p^2\v q^2}  f^{abc}\rho ^{*b}(-p-k)\rho ^d(-q)\alpha^{\dagger c}_j(k^+,-\v k)A^a_i(p^+,\v p)A^{\dagger d}_l(q^+,\v q)|0\rangle_F \nonumber\\
&&=i4g^3\int_{p^->k^-;\  p^+>k^+}\frac{1}{p^+}\frac{(\v p+\v k)_i}{\v p^2(\v p+\v k)^2}f^{abc}\rho ^a(p)\rho ^b(-p-k)\alpha^{\dagger c}_i(k)|0\rangle_F 
\end{eqnarray}
So that finally we get
\begin{eqnarray}\label{psinew}
\psi^a_{ \ soft}(k)&\equiv &\frac{1}{\sqrt{2k^+}}\langle \v k| \Psi_{soft}\rangle \nonumber \\
&=&\frac{i4g^3}{(2\pi)^{3/2}[2k^+]}\int_{p^->k^-;\  p^+>k^+}\frac{1}{p^+}\frac{(\v p+\v k)_i}{\v p^2(\v p+\v k)^2}f^{abc}\rho ^b(p)\rho ^c(-p-k)\nonumber\\
&=&\frac{i4g^3}{(2\pi)^{3/2}[2k^+]}\int_{p^->k^-;\  p^+>k^+}\frac{1}{p^+}\frac{(\v p-\v k)_i}{\v p^2(\v p-\v k)^2}f^{abc}\rho ^b(p-k)\rho ^c(-p)
\end{eqnarray}

\subsubsection*{Comparison with \cite{LuMu}}
Putting  together eq.(\ref{psi1}) and eq.(\ref{psinew}) we find
  \begin{equation}
\psi^a_{eik}(k)+\psi^a_{soft}(k)=  i\frac{4g^3}{(2\pi)^{3/2}[2k^+]}\left[2\int_{k^-<p^-; p+}-\int_{k^-<p^-;  p^+>k^+}\right]\frac{1}{p^+}\frac{1}{\v p^2}\frac{(\v p-\v k)_i}{(\v p-\v k)^2}f^{abc}\rho ^b(p-k)\rho ^c(-p)
  \end{equation}
  The integral over the longitudinal momentum can be easily performed and we obtain (for $\frac{E}{k^-}\gg \frac{\v p^2}{\v k^2}\gg 1$)
   \begin{equation}\label{psi1}
\psi^a_{eik}(k)+\psi^a_{soft}(k)=  i\frac{4g^3}{(2\pi)^{3/2}[2k^+]}\int {d^2\v p\over (2\pi)^2} \left[2\ln\frac{E}{k^-}-\ln\frac{\v p^2}{\v k^2}\right]\,\frac{1}{\v p^2}\frac{(\v p-\v k)_i}{(\v p-\v k)^2}f^{abc}\rho ^b(p-k)\rho ^c(-p)
  \end{equation}
This  reproduces the logarithmic contributions to the perturbative amplitude $\psi^3+\psi^4+\psi^5+\psi^6$ calculated in \cite{LuMu}. 
Subsequent integration over $k^-<E$ if interpreted in the same way as in \cite{LuMu} would then produce an NLO correction to the evolution kernel proportional to the transverse logarithm, as discussed earlier. We now perform this integration explicitly,
\begin{eqnarray}\label{psi2}
&&\int {dk^+\over 2\pi} [\psi^a_{eik}(k)+\psi^a_{soft}(k)]= 
{ i2g^3\over (2\pi)^{5/2}}\int \frac{dk^+}{k^+}\int {d^2\v p\over (2\pi)^2} \left[2\ln\frac{E}{k^-}-\ln\frac{\v p^2}{\v k^2}\right]\,\frac{1}{\v p^2}\frac{(\v p-\v k)_i}{(\v p-\v k)^2}f^{abc}\rho ^b(p-k)\rho ^c(-p)\nonumber \\
&&={ i2g^3\over (2\pi)^{5/2}}\int_E^{Ee^\Delta} \frac{dk^-}{k^-}\int {d^2\v p\over (2\pi)^2} \left[2\ln\frac{E}{k^-}-\ln\frac{\v p^2}{\v k^2}\right]\,\frac{1}{\v p^2}\frac{(\v p-\v k)_i}{(\v p-\v k)^2}f^{abc}\rho ^b(p-k)\rho ^c(-p)\nonumber \\
 &&={ i2g^3\over (2\pi)^{5/2}}\int {d^2\v p\over (2\pi)^2} \left[\Delta^2-\Delta\ln\frac{\v p^2}{\v k^2}\right]\,\frac{1}{\v p^2}\frac{(\v p-\v k)_i}{(\v p-\v k)^2}f^{abc}\rho ^b(p-k)\rho ^c(-p)
  \end{eqnarray}
This reproduces (\ref{sumpsi}) precisely with the same prefactors in \eqref{F2}.
  
We note here two interesting points. First, the gluon with momentum $k$ that is created in the wave function, in both the eikonal and soft contributions has the frequency which is smaller than the  frequency in the last rapidity bin. In this sense these contributions correspond to "back reaction" of the fastest modes on the slower modes, and not to production of the fast particles. 

The second point is that the transverse logarithm in expression \eqref{psi1} does not arise only from  imposition of double ordering in $p^->k^-$ and $p^+<k^+$. As discussed above this is indeed the nature of the eikonal contribution, where the transverse logarithm appears in the integral over $p^-$ due to the additional ordering of the longitudinal momenta which translates into $p^->k^-\frac{\v p^2}{\v k^2}$.
However the soft term is of a different nature. It provides an additional {\bf positive} contribution to the inclusive emission probability, logarithmic in the ratio of transverse momenta but independent on the total energy $E$ as long as $E/k^->\v p^2/\v k^2$. This is quite natural since this term produces gluons with frequencies far below $E$, and is independent of $E$ as long as both, the real gluon $k$ and the virtual gluon $p$ have frequencies much smaller than the energy.

We thus explicitly observe that the double ordering is {\bf NOT} the only origin of large transverse logarithms in the NLO calculation, but that part of it is due to emission of soft gluons.

This concludes our discussion of the correspondence between our present calculation and \cite{LuMu}.

  \section{Conclusions}
  In this work we have developed the Born Oppenheimer renormalization group approach to the QCD evolution. The physical evolution parameter is taken to be the frequency ($k^-$) of the gluon field modes.  The modes relevant to scattering at any point of the evolution are  those with frequencies smaller than the frequency cutoff. Increasing this cutoff requires including  additional field modes in the wave function. Since the new modes have high frequency, the calculation of their wave function is performed in the Born-Oppenheimer approximation, i.e. the wave function of the fast modes is calculated in the fixed background of the slow field modes. The process is iterated as the frequency cutoff is increased. 
  
  This approach unifies the high energy evolution, in which case the increase in frequency is associated with the increase of the energy of scattering, and the perturbative evolution in $Q^2$ where the frequency increases due to increase in transverse resolution. Importantly, the evolution of the hadronic wave function is due to both types of splittings: the BFKL type where the emitted gluon has the smallest longitudinal momentum, and the DGLAP type where the transverse momentum increases in splitting with or without significant change of the longitudinal momentum. Thus even if we are interested in evolution in energy, the DGLAP type splittings in the wave function are still present and in principle affect the evolution of the eikonal scattering cross section with the target.  We note that DGLAP splittings are present in JIMWLK evolution as well, however there they appear as NLO corrections in the JIMWLK kernel, and need to be resummed separately \cite{ourdglap}. The effects of the evolution will be discussed in detail in subsequent papers.
  
  In this paper we set out the general framework of the approach and prepare ground for calculations of evolution. This includes first of all deriving the interaction Hamiltonian between the fast and slow modes. In accordance with the previous discussion, we explicitly demonstrate that the interaction contains the BFKL type eikonal vertices as well as DGLAP type vertices with strong ordering of transverse momentum. Both types are important for the evolution and both are integral part of the Born Oppenheimer approximation. Using this interaction Hamiltonian we derive perturbatively the wave function of a hadronic system evolved from some initial frequency $E_0$ to a final frequency $E$. 
  We also derive the form of the $S$-matrix consistent with our approximations, similarly to how the eikonal $S$-matrix is consistent with the eikonal approximation for the evolution in the standard BFKL approach. Here we limit ourselves to scattering on dilute target, and thus find the scattering matrix in second order in the target fields.
  
  Finally, we demonstrate that expansion of the wave function we have derived here to fixed order in the QCD coupling constant reproduces the calculation of \cite{LuMu} of NLO JIMWLK Hamiltonian. We identify the large transverse logarithms observed in \cite{LuMu} with two types of effects. The first one is the fact that not all emissions of fast modes in the evolution are eikonal, but only those in which the longitudinal momentum is strongly ordered. The second effect is the emission of soft gluons with momenta (both longitudinal and transverse) and frequencies much smaller than the rest of gluons in the interaction vertex.
 
 In the companion paper and subsequent work we consider in detail BO evolution of physical observables. 
 
 In \cite{second} we show that the evolution of the gluon TMD is given, in the linear approximation by the CSS equation and discuss in detail the relation between the frequency evolution parameter in BO on one hand, and the transverse and longitudinal resolution scales of CSS on the other. Beyond linear approximation the BO evolution also yields nonlinear corrections which have a simple intuitive interpretation as stimulated emission corrections due to nonvanishing gluon density. At high $Q^2$ these corrections are suppressed by a factor $1/Q^2R^2$ with $R^2$ - the area of the hadron. This is the same suppression factor as in the nonlinear GLR corrections, however the physical origin and parametric dependence of the stimulated emission and GLR terms are very different. The stimulated emission correction is $O(\alpha_s)$, while the GLR term is $O(\alpha_s^2)$. On the other hand the GLR term is enhanced at low $x$ by the factor $\ln 1/x$, while the stimulated emission is not enhanced at low $x$.
 
 We also consider the evolution of gluon PDF. Here we show that at not too small $x$ it is governed by the DGLAP equation with a subleading twist term due to the stimulated emission. Interestingly, the stimulated emission term associated with virtual corrections is larger than the one associated with real correction, and the net effect is a slowdown of DGLAP evolution. The effect has therefore the same sign as the GLR shadowing corrections, although, as mentioned above the two have very different physical origin.
 
  The analysis of the scattering matrix in the soft scattering regime and its relation to the BFKL equation is left to subsequent work which will appear soon \cite{third}.
  The next natural step which is left for future work, is the extension of this approach to include multiple scattering and saturation effects.
  
  \section*{Acknowledgments}

 The research was supported by  the  Binational Science Foundation grant \#2012124;  MSCA RISE 823947 “Heavy ion collisions: collectivity and precision in saturation physics” (HIEIC), and by VATAT (Israel planning and budgeting committee) grant for supporting
theoretical high energy physics.
 The work of H.D. and A.K. is supported by the NSF Nuclear Theory grant \#2208387.
The work of ML was  funded by Binational Science Foundation grant \#2021789  and by the ISF grant \#910/23.\\
We  thank Physics Departments of the Ben Gurion University and University of Connecticut for hospitality during mutual visits. \\
We also thank EIC Theory Institute, Brookhaven National Laboratory, CERN-TH group, 
National Centre for Nuclear Research, Warsaw;  Institute for Nuclear Theory at  the University of Washington, ECT$^*$,
 and ITP at the University of Heidelberg  for their support and hospitality during various stages of completing this project.

\appendix

\section{More on $\psi_i^n$}\label{psisec}

In this appendix we give some details of algebraic manipulations leading to expressions quoted in Sec.4.

As in \cite{LuMu}, we split the contribution to $\psi^3_i$ into the up and down
\begin{equation}
\psi^{3u}_i=\theta(p^+-k^+)\psi^3_i;\ \ \ \ \ \psi^{3d}_i=\theta(k^+-p^+)\psi^3_i
\end{equation}
After a little algebra we find
\begin{eqnarray}
\psi^4_i&=&-\frac{k^+-p^+}{(k^++p^+)^2}\frac{k^+}{p^+}\frac{\v p_i}{\v k^2[k^+p^2+p^+\v k^2]}\\
\psi^5_i+\psi^{3u}_i&=&\theta(p^+-k^+)\frac{1}{\v k^2\v p^2}\frac{k^+}{p^+}\frac{1}{[k^+(\v k-\v p)^2+(p^+-k^+)\v k^2]}\times\nonumber\\
&&\Bigg\{\Big[-\frac{p^+\v k^2+k^+\v p^2}{k^+(k^+-p^+)}-2\frac{1}{p^+}\v p^2\Big]\v p_i+2\Big[\frac{\v p^2}{p^+}-\frac{\v p\cdot (\v k-\v p)}{k^+}\Big]\v k_i\Bigg\}\\
\psi^6_i+\psi^{3d}_i&=&\theta(k^+-p^+)\frac{1}{\v k^2\v p^2}\frac{1}{[p^+(\v k-\v p)^2+(k^+-p^+)\v p^2]}\times\nonumber\\
&&\Bigg\{\Big[\frac{k^+\v p^2+p^+\v k^2}{p^+(k^+-p^+)}-2\frac{1}{p^+}\v p^2\Big]\v p_i+2\Big[\frac{\v p^2}{p^+}-\frac{\v p\cdot (\v k-\v p)}{k^+}\Big]\v k_i\Bigg\}
\end{eqnarray}
Anticipating that these expressions will be integrated  over the longitudinal momenta and transverse momentum $\v p$,
we  change $p^+\rightarrow p^++k^+$ in $\psi^5$ and $\psi^{3u}$. This eliminates the explicit $\theta$-function in 
these two terms.
Additionally we make the transformation $\v p_i\rightarrow \v k_i-\v p_i$. This transformation has an additional effect that $F^2\rightarrow -F^2$, so we have to account for that extra sign. 
We get
\begin{eqnarray}
\psi^5_i&=&\frac{1}{\v k^2(\v p-\v k)^2}\frac{1}{p^++k^+}\frac{1}{p^+}\frac{k^+\v p^2}{[k^+\v p^2+p^+\v k^2]}\times\\
&&\Bigg\{2\Big[-\frac{k^+}{p^+}  -\frac{\v p\cdot \v k}{\v p^2}-\frac{(\v k-\v p)^2}{\v p^2}\frac{p^+}{k^++p^+}
\Big]\v p_i+2\Big[\frac{k^+}{p^+}+\frac{\v p\cdot (\v k-\v p)}{\v p^2}\frac{p^+}{k^+}+\frac{\v p\cdot\v k}{p^2}\Big]\v k_i\Bigg\}\nonumber\\
\psi^{3u}_i&=&-\frac{2k^++p^+}{(p^+)^2}\frac{1}{p^++k^+}\frac{1}{\v k^2(\v p-\v k)^2}(\v k_i-\v p_i)\nonumber \end{eqnarray}


After a little more algebra we can rewrite the terms in the following way
\begin{eqnarray}
\psi^5_i&=&\frac{2k^++p^+}{(p^+)^2(p^++k^+)}\frac{1}{\v k^2}\frac{k^+p^2}{k^+\v p^2+p^+\v k^2}\frac{(\v k-\v p)_i}{(\v k-\v p)^2}
+\frac{k^+-p^+}{(k^++p^+)^2p^+}\frac{1}{\v k^2}\frac{k^+\v p^2}{(k^+\v p^2+p^+\v k^2)}\frac{\v p_i}{\v p^2}\nonumber\\
&+&\frac{1}{(p^++k^+)p^+}\frac{k^+}{k^+\v p^2+p^+\v k^2}\left[\frac{(\v k-\v p)_i}{(\v k-\v p)^2}-\frac{\v k_i}{\v k^2}\right]
+2\frac{1}{(p^++k^+)(k^+\v p^2+p^+\v k^2)}\frac{\v p\cdot(\v k-\v p)}{ (\v k-\v p)^2}\frac{\v k_i}{\v k^2}\\
\psi^{3u}_{i}&=&-\frac{2k^++p^+}{(p^+)^2(p^++k^+)}\frac{1}{\v k^2}\frac{\v (\v k-\v p)_i}{(\v k-\v p)^2}\nonumber\\
\psi^4_i&=&-\frac{k^+-p^+}{(k^++p^+)^2p^+}\frac{1}{\v k^2}\frac{k^+\v p^2}{(k^+\v p^2+p^+\v k^2)}\frac{\v p_i}{\v p^2}
\end{eqnarray}
Some more trivial algebra gives
\begin{eqnarray}
\psi^5_i&=&-\psi^{3u}_i-\psi^4_i-\left(\frac{2}{p^+}-\frac{1}{p^++k^+}\right)\frac{1}{k^+\v p^2+p^+\v k^2}\frac{(\v k-\v p)_i}{(\v k-\v p)^2}  \\
&+&
\left(\frac{1}{p^+}-\frac{1}{p^++k^+}\right)\frac{1}{k^+\v p^2+p^+\v k^2}\left[\frac{(\v k-\v p)_i}{(\v k-\v p)^2}-\frac{\v k_i}{\v k^2}\right]
+\frac{2}{p^++k^+}\frac{1}{k^+\v p^2+p^+\v k^2}\frac{\v p\cdot(\v k-\v p)}{(\v k-\v p)^2}\frac{\v k_i}{\v k^2}\nonumber\\
&=&-\psi^{3u}_i-\psi^4_i\nonumber\\
&-&\frac{1}{p^+}\frac{1}{k^+\v p^2+p^+\v k^2}\frac{(\v k-\v p)_i}{(\v k-\v p)^2}-\left(\frac{1}{p^+}-\frac{1}{p^++k^+}\right)\frac{1}{k^+\v p^2+p^+\v k^2}\frac{\v k_i}{\v k^2}+\frac{2}{p^++k^+}\frac{1}{k^+\v p^2+p^+\v k^2}\frac{\v p\cdot(\v k-\v p)}{(\v k-\v p)^2}\frac{\v k_i}{\v k^2}\nonumber
\end{eqnarray}
In addition we have
\begin{equation}
\psi^6_i+\psi^{3d}_i
=\theta(k^+-p^+)\frac{1}{\v k^2\v p^2}\frac{1}{[p^+(\v k-\v p)^2+(k^+-p^+)\v p^2]}\Bigg\{\Big[\frac{\v p^2+\v k^2}{k^+-p^+}-\frac{1}{p^+}\v p^2\Big]\v p_i+2\Big[\frac{\v p^2}{p^+}-\frac{\v p\cdot(\v k-\v p)}{k^+}\Big]\v k_i\Bigg\}
\end{equation} 
The amplitude $\psi^6_i$ contains no large logarithms, since there is no large transverse momentum hierarchy between incoming and outgoing gluons. It is nevertheless important since it contributes ${\Delta}^2$ terms upon integration over the longitudinal momenta. We can rewrite this contribution  by changing $p^+\rightarrow k^+-p^+,\ \v p\rightarrow \v k-\v p$ in the terms that have the denominator different from $1/p^+$. The result is
\begin{equation}\label{63}
\psi^6_i+\psi^{3d}_i
=\theta(k^+-p^+)\frac{1}{\v k^2(\v p-\v k)^2}\frac{1}{[p^+(\v k-\v p)^2+(k^+-p^+)\v p^2]}\left[(\v k^2\v p_i-\v p\cdot\v k \v k_i)\frac{1}{p^+}+\v p\cdot(\v p-\v k)\v k_i\left[\frac{1}{p^+}-\frac{2}{k^+}\right]\right]
\end{equation}
The first term contributes ${\Delta}^2$ upon integration over $p^+$ and $k^+$. The second although formally also does, in fact does not contain such double divergence, because of the symmetry $p\rightarrow k-p$, since the double divergence comes from the region $p^+\ll k^+$, and in this regime the energy denominator becomes $\frac{1}{\v p^2k^+}$.

A useful identity
\begin{equation}
\frac{(\v k-\v p)_i}{(\v k-\v p)^2}-\frac{\v k_i}{\v k^2}=-\frac{\v k^2\v p_i-\v p\cdot\v k \v k_i}{\v k^2(\v k-\v p)^2}+\frac{\v p\cdot(\v k-\v p)}{(\v k-\v p)^2}\frac{\v k_i}{\v k^2}
\end{equation}
Combining all the terms together (while omitting the last term in eq.(\ref{63}) which clearly does not lead to a logarithmic contribution)) we arrive at \eq{psiABC}.

\section{More on the soft modes $\alpha_i^a$}\label{soft}

In the main body of this paper we did not treat too carefully the interaction involving the soft modes $\alpha$, i.e. $\Delta {\cal H}$ (\ref{dH}). We justified its omission to first order in $g$ when calculating the evolution of the wave function. However when comparing with the results of \cite{LuMu} we omitted it even though there the wave function on the rapidity extended interval $\Delta\gg 1$ was needed to $O(g^3)$.
In this appendix we show that this additional interaction indeed cancels.

The wave function we used in Sec. 4 is generated by the coherent operator that creates the classical solution of the equation of motion with  $\Delta {\cal H}$ (\ref{dH}) excluded. Our goal here is precisely to show that excluding the contribution of  $\Delta {\cal H}$ (\ref{dH}) from the equation of motion is justified due to a subtle cancellation of this term with the effects of noncommutativity of the color charge density operators $\rho^a$.
Let us show this explicitly.

The classical equation of motion for $A_j^b(p)$ in the background $\alpha$ is
\begin{eqnarray}
\frac{g}{2p^+}\v P^{ab}_i\rho^b(-p)&+&\frac{1}{2}\left[\v P^2\delta_{ij}+[\v P_i,\v P_j]\right]^{ab}A^b_j(p)\nonumber \\
&+&igp^+f^{abc}\int_{k^+\ll p^+; k^-\ll p^-}\frac{\v k_j}{k^+}\left[\alpha_j^{ b}(k^+,\v k)-\alpha_j^{\dagger b}(k^+,-\v k)\right]A^c_i(p)
=0
\end{eqnarray}
The last term here arises from the  $\Delta {\cal H}$ (\ref{dH}), which now we keep in full glory.
Perturbative solution of this equation to order $g^2$ is
\begin{eqnarray}\label{clas}
A_i^a(p^+,\v p)&=&-g\frac{1}{p^+}\frac{\v p_i}{\v p^2}\rho^a(-p)-ig^2\frac{1}{p^+}\frac{1}{\v p^2}f^{abc}\int_{k^+\ll p^+; k^-\ll p^-}\left[\alpha_i^{\dagger c}(k^+,\v k)+\alpha_i^{ c}(k^+,-\v k)\right]\rho^b(-p)\\
&-&2ig^2\frac{\v p_i}{\v p^4}\frac{1}{p^+}f^{abc}\int_{k^+\ll p^+; k^-\ll p^-}\left[ \v p_j-\frac{p^+}{k^+}\v k_j\right]\left[\alpha_j^{\dagger b}(k^+,\v k)+\alpha_j^{ b}(k^+,\v k)\right]\rho^c(-p)=-g\frac{1}{p^+}\frac{\v p_i}{\v p^2}\rho^a(-p)\nonumber\\
&+&ig^2\frac{1}{\v p^2}\frac{1}{p^+}f^{abc}\int_{k^+\ll p^+; k^-\ll p^-}\left\{\left[\delta_{ij}- \frac{2\v p_i\v p_j}{\v p^2}\right]+\frac{2p^+}{k^+}\frac{\v p_i\v k_j}{\v p^2}\right\}\left[\alpha_j^{\dagger b}(k^+,\v k)+\alpha_j^{ b}(k^+,\v k)\right]\rho^c(-p)\nonumber 
\end{eqnarray}
It is interesting to note that the $\delta_{ij}$ term on the RHS originates from the covariant derivative acting on $\rho$ and corresponds precisely to $\psi^3_{gg}$ of \cite{LuMu} (the two gluon production amplitude that we did not explicitly quote in this text), while the $\v p_i\v p_j$ term corresponds to the penultimate term in $\psi^2_{gg}$.
The last term in this expression corresponds to the last term in the amplitude $\psi^2_{gg}$ in \cite{LuMu}. However we also know from \cite{LuMu} that in the final expression it is cancelled by the appropriate term in $\psi^1_{gg}$. In our calculation  such a cancellation occurs as well, as we demonstrate below. The final upshot is that the relevant perturbative expression for the field $A^a_i(p)$ is
\begin{equation}
A_i^a(p^+,\v p)=-g\frac{1}{p^+}\frac{\v p_i}{\v p^2}\rho^a(-\v p)
+ig^2\frac{1}{\v p^2}\left[\delta_{ij}- \frac{2\v p_i\v p_j}{\v p^2}\right]\frac{1}{p^+}f^{abc}\int_{k^+\ll p^+; k^-\ll p^-}\left[\alpha_j^{\dagger b}(k^+,\v k)+\alpha_j^{ b}(k^+,\v k)\right]\rho^c(-\v p)
\end{equation}
This expression  corresponds to (\ref{wfun1}).

The coherent operator entering (\ref{wfun1}) can  be derived in the following way. 
First, consider a simplified toy Hamiltonian where $A$ and $\alpha$ interact only with $\rho$
\begin{equation}\label{h}
h=p^-A_i^{a\dagger} A_i^a+\rho^a(p)A^{a-}(p)+k^-\alpha^{a\dagger}_i\alpha^a_i+\rho^a(k)\alpha^{a-}(k)
\end{equation}
where $A^{a-}(p)=\frac{\v p_i}{p^+}A^{a}_i(p)$, and $\alpha^{a-}(k)=\frac{\v k_i}{k^+}\alpha^a_i(k)$
Now, suppose for a moment that we are not applying the Born-Oppenheimer formalism, but instead are simply solving the theory perturbatively. That is to leading order we diagonalize the kinetic term for $A$ and $\alpha$. Then the evolution operator to infinite time, which is also the operator that diagonalizes the Hamiltonian, can be written as 
(below $t$ means $x^+$)
\begin{equation}
U={\cal P}e^{i\int_0^\infty dt \left[\rho^a(p)A^{a-}(t,p)+\rho^a(k)\alpha^{a-}(t,k)\right]^I}
\end{equation}
where the superscript $I$ indicates the operator in the interaction representation. The operator should be understood as defined in such a way that the contribution at $\infty$ is killed by an appropriate $i\epsilon$ prescription.
Now expanding $U$  to first order we obtain (we are being sloppy with unimportant factors).
\begin{eqnarray}\label{u}
U&=&1+i\int_0^\infty dt \left[\rho^a(p)A^{a-}(p)e^{-ip^-t}+\rho^a(k)\alpha^{a-}(k)e^{-ik^-t}\right]=1+\left[\frac{1}{p^-}\rho^a(p)A^{a-}(p)+\frac{1}{k^-}\rho^a(k)\alpha^{a-}(k)\right]\nonumber\\
&=&1+i\left[\frac{\v p_i}{\v p^2}\rho^a(p)A_i^{a}(p)+\frac{\v k_i}{\v k^2}\rho^a(k)\alpha_i^{a}(k)\right]\approx e^{i\left[\frac{\v p_i}{\v p^2}\rho^a(p)A_i^{a}(p)+\frac{\v k_i}{\v k^2}\rho^a(k)\alpha_i^{a}(k)\right]}
\end{eqnarray}
where in the last line we have exponentiated the leading order result.
This is precisely the coherent operator we have been always using.  

However there is a catch here. The exponentiation of the linear term is fine if the coefficients in the Hamiltonian are $c$-numers. However in our case those involve the color charge operators which do not commute. We therefore have to reexamine the question of exponentiation.
Let us expand  consider the second order contribution to $U$  to check whether it is reproduced by our exponentiation. We are only interested in terms linear in $A$, since $A$ is supposed to be small
\begin{eqnarray}\label{u2}
&&U_2= \\
&&-\int_0^\infty dt_1 \left[\rho^a(p)A^{a-}(p)e^{-ip^-t_1}+\rho^a(k)\alpha^{a-}(k)e^{-ik^-t_1}\right]\int_0^{t_1} dt_2\left[\rho^a(p)A^{a-}(p)e^{-ip^-t_2}+\rho^a(k)\alpha^{a-}(k)e^{-ik^-t_2}\right]\nonumber\\
&&\approx-\int_0^\infty dt_1\int_0^{t_1} dt_2 \left[\rho^a(p)A^{a-}(p)e^{-ip^-t_1}\rho^b(k)\alpha^{b-}(k)e^{-ik^-t_2}+\rho^a(k)\alpha^{a-}(k)e^{-ik^-t_1}\rho^b(p)A^{b-}(p)e^{-ip^-t_2}\right]\nonumber \\
&&=-\frac{1}{2}\int_0^\infty dt_1\int_0^\infty dt_2 \left[\rho^a(p)A^{a-}(p)\rho^b(k)\alpha^{b-}(k)e^{-ip^-t_1-ik^-t_2}+\rho^a(k)\alpha^{a-}(k)\rho^b(p)A^{b-}(p)e^{-ik^-t_1-ip^-t_2}\right]\nonumber\\
&&+\frac{1}{2}\int_0^\infty dt_1\int_{t_1}^\infty dt_2  \left([\rho^a(p)A^{a-}(p),\rho^b(k)\alpha^{b-}(k)]e^{-ip^-t_1-ik^-t_2}+[\rho^a(k)\alpha^{a-}(k),\rho^b(p)A^{b-}(p)]e^{-ik^-t_1-ip^-t_2}\right)\nonumber
\end{eqnarray}
The first term is precisely what we have already accounted for in exponentiation in eq.(\ref{u}), while the second term is not contained in expansion of \eqref{u}. It therefore has to be calculated separately and added to the exponent in order to have a correct evolution operator.
We now assume that $k\ll p$, $k^-\ll p^-$ and $k^+\ll p^+$. Since $A$ and $\alpha$ commute, the only contribution to the commutator in \eqref{u2} comes from the commutator of the charge densities
\begin{eqnarray}
\delta U_2&=&\frac{1}{2}if^{abc}\rho^c(p)A^{b-}(p)\alpha^{c-}(k)\int_0^\infty dt_1\int_{t_1}^\infty dt_2 \{e^{-ip^-t_1}e^{-ik^-t_2}-
e^{-ik^-t_1}e^{-ip^-t_2}\}\nonumber\\
&=&\frac{1}{2}if^{abc}\rho^c(p)A^{b-}(p)\alpha^{c-}(k)\left[\frac{1}{p^-k^-}-\frac{1}{p^-p^-}\right]
=\frac{1}{2}if^{abc}\rho^c(p)\frac{\v p_i\v k_j}{\v p^2\v k^2} \left[1-\frac{\v k^2}{\v p^2}\frac{p^+}{k^+}\right]A_i^{b}(p)\alpha_j^{c}(k)
\end{eqnarray}
This expression precisely contains the contribution of the two diagrams that contribute to $\psi^1_{gg}$ of \cite{LuMu}. 
Adding it to the exponent in \eqref{u}  we arrive to
\begin{equation}\label{expo}
U=\exp\left\{i\left[\frac{\v p_i}{\v p^2}\rho^a(p)A_i^{a}(p)+\frac{\v k_i}{\v k^2}\rho^a(k)\alpha_i^{a}(k)\right] +\frac{1}{2}if^{abc}\rho^c(p)\frac{\v p_i\v k_j}{\v p^2\v k^2} \left[1-\frac{\v k^2}{\v p^2}\frac{p^+}{k^+}\right]A_i^{b}(p)\alpha_j^{c}(k)\right\}
\end{equation}
We notice that the origin of the extra term here is solely due to non-commutativity of $\rho$.
In other words, linear interaction is exactly diagonalized by coherent operator unless the background is non-commutative. For noncommutative background we need to add the extra term.

In our calculation in Sec. 4, we are not doing perturbation theory in the $\alpha$ sector. To understand what happens then, let us assume that we have solved completely for the wave function of $\alpha$ and $\rho$, and are only calculating the evolution operator for $A$. The interaction picture therefore should be defined with $H_0$ which contains the complete dynamics of $\alpha$ and $\rho$, and the "perturbative" time evolution of $H_A$ should be given by this Hamiltonian. The crucial element here is that in this case $\rho$ cannot b considered to be static any more. For $H_\alpha$ given by the last two terms in eq.(\ref{h}) the time evolution of $\rho$ is given by the usual covariant conservation equation
\begin{equation}
\frac{d}{dt}\rho^a(p)=-if^{abc}\alpha^{-b}(k)\rho^c(p)
\end{equation}
for which the solution is
\begin{equation}\label{rt}
\rho^a(p,t)=\left[{\cal P}\exp\{-i\int_{0}^t d\tau T\alpha^-(\tau)\}\right]^{ab}\rho^b(p)
\end{equation}
Assuming for simplicity  that  $\alpha^-$ is time independent, since by definition its frequency is the lowest one around,  we have
\begin{equation}\label{rhoe}
\rho^a(p,t)\approx \left[\exp\{-i T\alpha^-t\}\right]^{ab}\rho^b(p)
\end{equation}
We can now calculate the evolution operator (diagonalizing operator ) for the $A$ field using the first two terms of eq.(\ref{h}) and the interaction picture time dependence of $\rho$ from eq.(\ref{rhoe}).
\begin{eqnarray}
U(A)&=&1-i\int_0^\infty dt \rho^a(p,t)A^{a-}(p,t)=1+i\rho^a(p)\int_0^\infty dt\left[ e^{-i\left(-T\alpha^-+p^-\right)t}\right]_{ab}A^{-b}\nonumber\\
&=&1+\rho^a(p)\left[\frac{1}{p^--T\alpha^-}\right]_{ab}A^{-b}\approx\exp\left\{i\frac{p_i}{p^+}\rho^a(p)\left[\frac{1}{p^--T\alpha^-}\right]_{ab}A_i^b(p)\right\}
\end{eqnarray}
Expanding the last expression to first order in $\alpha$ we recover the first and last terms in eq. (\ref{expo}). The other two terms are recovered once we remember that the total evolution operator contains the product $U(A)U(\alpha)$. Keeping only the usual "coherent operator" in $U(\alpha)$ to first order in $\alpha$ we recover also the rest of the terms in eq.(\ref{expo}).

We are now ready to tackle the main problem, that is to understand the effect of $\Delta {\cal H}$ (\ref{dH}) on the wave-function of $A$. We follow exactly the same strategy as above. First, consider all quadratic in $A$ terms. Those define the propagator, or in the context of our calculation the frequency. Define
\begin{equation}
P^-\equiv \frac{\v P^2}{2p^+}
\end{equation}
In terms of this, the frequency of the field $A$, i.e. the time dependence of the Heisenberg operator $A$ is given by 
\begin{equation}
A^a_i(p,t)=\left[e^{-i(P^-+T\alpha^-)t}\right]_{ab}A^b_i(p)
\end{equation}
The charge density $\rho$ has the same time dependence as in eq.(\ref{rt}). Now the diagonalizing operator is
\begin{eqnarray}\label{ua}
U(A)&=&1-i\int_0^\infty dt \rho^a(p,t)A^{a-}(p,t)=1+i\rho^a(p)\int_0^\infty dt\left[ e^{iTa^-t}e^{-i(P^-+T\alpha^-)t}\right]_{ab}A^{-b}(p)\nonumber\\
&=&1+i\rho^a(p)\left[\frac{\v P_i}{\v P^2}\right]_{ab}A_i^{b}(p)\approx e^{i\rho^a(p)\left[\frac{\v P_i}{\v P^2}\right]_{ab}A_i^{b}(p)}
\end{eqnarray}
Note that in the first line the dependence on $\alpha^-$ in the phase exactly cancels between the time dependence of $A$ and the time dependence of $\rho$, which corresponds precisely to the cancellation of appropriate terms between $\psi^1_{gg}$ and $\psi^2_{gg}$ in \cite{LuMu}. 

The end result therefore is that the effect of the extra interaction Hamiltonian  $\Delta {\cal H}$ (\ref{dH}) is nagated completely by the fact that the color charge density operators are noncommuting. As a result, when defining the "classical" field which enters the coherent operator we are allowed to forget about  $\Delta {\cal H}$ (\ref{dH}) and simultaneously treat $\rho$ naively as commuting, when exponentiating the leading order solution.




\bibliographystyle{unsrt}
\bibliography{BORGpaper1}
\end{document}